\documentclass[twocolumn,tighten]{aastex6}

\usepackage{multirow}

\usepackage{subfigure}
\usepackage{amsmath}
\usepackage{mathtools}
\usepackage[all]{hypcap}

\newcommand{\msun}{\ensuremath{M_{\odot}}}
\newcommand{\kmpers}{\ensuremath{\mathrm{km\,s}^{-1}}}

\newcommand{\diff}{\textsl{d}}
\newcommand{\vect}[1]{\boldsymbol{#1}}

\definecolor{pymagenta}{rgb}{0.75,0,0.75}
\definecolor{pyblue}{rgb}{0,0,1}
\definecolor{pygreen}{rgb}{0,0.5,0}
\definecolor{pyred}{rgb}{1,0,0}
\definecolor{pycyan}{rgb}{0,0.75,0.75}

\begin{document}

\title{Hydrodynamical neutron-star kicks in electron-capture supernovae and implications for
the CRAB supernova}

\shorttitle{Neutron-star kicks in electron-capture supernovae and the Crab supernova}
\shortauthors{A.~Ge{\ss}ner and H.-Th.~Janka}
\author{Alexandra~Ge{\ss}ner\altaffilmark{1,2} and Hans-Thomas~Janka\altaffilmark{2}}
\altaffiliation{E-mail: alexandra.gessner@tuebingen.mpg.de}
\altaffiliation{E-mail: thj@mpa-garching.mpg.de}
\affil{\altaffilmark{1} Max Planck Institute for Intelligent Systems, Max-Planck-Ring~4, 
       72076 T\"ubingen, Germany\\
\altaffilmark{2} Max Planck Institute for Astrophysics, 
       Karl-Schwarzschild-Str.~1, 85748 Garching, Germany}

\begin{abstract}
Neutron stars (NSs) obtain kicks of typically several 100\,km\,s$^{-1}$ at birth. The gravitational
tug-boat mechanism can explain these kicks as consequences of asymmetric mass ejection during the
supernova (SN) explosion. Support for this hydrodynamic explanation is provided by observations of
SN remnants with associated NSs, which confirm the prediction that the bulk of the 
explosion ejecta, in particular chemical elements between silicon and the iron group,
are dominantly expelled in the hemisphere opposite to the direction of the NS kick.
Here, we present a large set of two- and three-dimensional explosion simulations of
electron-capture SNe, considering explosion energies between
$\sim$3$\times 10^{49}$\,erg and $\sim$1.6$\times 10^{50}$\,erg. We find that the
fast acceleration of the SN shock in the steep density gradient delimiting the O-Ne-Mg
core of the progenitor enables such a rapid expansion of neutrino-heated matter that the
growth of neutrino-driven convection freezes out quickly in a high-mode spherical 
harmonics pattern. Since the corresponding momentum asymmetry of the ejecta is very small
and the gravitational acceleration by the fast-expanding ejecta abates rapidly,
the NS kick velocities are at most a few km\,s$^{-1}$. The extremely low core compactness
of O-Ne-Mg-core progenitors therefore favors hydrodynamic NS kicks much below the 
$\sim$160\,km\,s$^{-1}$ measured for the Crab pulsar. This suggests either that the Crab
Nebula is not the remnant of an electron-capture SN, but of a low-mass iron-core progenitor,
or that the Crab pulsar was not accelerated by the gravitational tug-boat mechanism but 
received its kick by a non-hydrodynamic mechanism such as, e.g., anisotropic neutrino emission.
\end{abstract}


\section{Introduction}
\label{sec:intro}

Observations of young radio pulsars and measured orbital parameters of 
neutron stars (NSs) in close binary systems yield evidence that NSs acquire average
kick velocities between 200\,km\,s$^{-1}$ and 500\,km\,s$^{-1}$
during their birth in supernova (SN) explosions
\citep[e.g.,][]{Helfandetal1977,Tutukovetal1984,Harrisonetal1993,Kaspietal1996,
LyneLorimer1994,HansenPhinney1997,CordesChernoff1998,
Fryeretal1998,Laietal2001,Arzoumanianetal2002,Chatterjeeetal2005,Hobbsetal2005}.
The most likely explanation of these natal kick velocities is a recoil of the
newly formed NS due to anisotropic mass ejection when the core-collapse SN
explosion is initiated in a nonspherical way. The corresponding acceleration
of the NS can be understood by the ``gravitational tug-boat mechanism'', in
which the asymmetrically expelled matter exerts a long-lasting momentum transfer
to the NS by hydrodynamic and gravitational forces over a period of many 
seconds \citep{Schecketal2004,Schecketal2006,Nordhausetal2010,Nordhausetal2012,
Wongwathanaratetal2010,Wongwathanaratetal2013,Bruennetal2016,Janka2017,Muelleretal2017}.

This hydrodynamic kick scenario is supported by recent observational analyses of SN 
remnants with associated NSs \citep{Holland-Ashfordetal2017,Katsudaetal2018,BearSoker2017}, 
which confirm the prediction by the gravitational tug-boat mechanism that the
bulk of the SN ejecta should be found moving in the hemisphere opposite to the
direction of the NS kick velocity \citep{Wongwathanaratetal2013,Wongwathanaratetal2017}. 
This anti-alignment is imprinted in the early moments of the explosion
on the chemical yields (from Si to the Fe-group) that are produced in the immediate
surroundings of the new-born NS and thus are most strongly affected by anisotropic
shock propagation. The asymmetry becomes more prominent with higher NS kicks.
Aging SN remnants, however, will develop a global hemispheric asymmetry due to
the changing relative motion of the NS and the ejecta. The 
SN ejecta are gradually decelerated by swept-up
circumstellar material while the NS moves within the gaseous ejecta cloud
with essentially maintained speed. Discriminating such an apparent kick-ejecta
anti-orientation and possible effects due to environmental asymmetries from
intrinsic explosion and ejecta asymmetries becomes an increasingly challenging
task for older SN remnants \citep{Katsudaetal2018}.
 
Two-dimensional (axisymmetric; 2D) models \citep{Schecketal2004,Schecketal2006,
Nordhausetal2010,Nordhausetal2012,Bruennetal2016}
as well as three-dimensional (3D) simulations 
\citep{Wongwathanaratetal2010,Wongwathanaratetal2013,Muelleretal2017} 
for iron-core progenitors of 15--20\,$M_\odot$ have shown that
the gravitational tug-boat mechanism can well explain NS kicks up to more than
1000\,km\,s$^{-1}$, because considerable dipolar asymmetry in the SN ejecta
can result from violent mass motions by convective overturn in the 
neutrino-heating layer, from large-amplitude sloshing and spiral activity of the
stalled SN shock due to the standing accretion-shock instability 
\citep[SASI; e.g.,][]{Blondinetal2003,Foglizzo2002,Foglizzoetal2006,
Foglizzoetal2007,Schecketal2008} or from long-lasting,
anisotropic accretion downdrafts around the nascent NS after the onset of the
SN blast \citep{Muelleretal2017}. 
Depending on stochastic variations of such large-scale
asymmetry, the kick velocities were found to range from less than 
10\,km\,s$^{-1}$ to over 1000\,km\,s$^{-1}$, with average values of
several 100\,km\,s$^{-1}$. 

Besides such stochastic case-to-case variations, hydrodynamic NS kicks associated
with SN ejecta asymmetries are expected to exhibit systematic dependences on
the explosion and progenitor properties. On average, larger kicks 
should be the result of higher explosion energies and of higher masses of the
anisotropically ejected matter that accelerates the NS by momentum transfer
\citep{Janka2017}. The kick-relevant mass involved in this process 
depends on the ``compactness'', $\xi_M = (M/M_\odot)/(R(M)/1000\,\mathrm{km})$,
of the progenitor \citep[defined in][]{OConnorOtt2011},
which is correlated with the steepness
of the density profile around the degenerate stellar core at the onset of
collapse \citep{SukhboldWoosley2014}.
Progenitors with shallow density profiles possess high compactness
values. If they explode by the neutrino-driven mechanism, a lot of matter is
available to absorb energy from neutrinos, allowing for high explosion energies
and large amounts of kick-mediating ejecta. This relation
explains the trend to higher NS kicks (on average) for successfully exploding
high-compactness progenitors \citep{Janka2017}.
The opposite tendency should apply for progenitors with steep density profiles
and correspondingly lower values of the core compactness. Indeed, a small set
of 2D and 3D explosion simulations by \citet{Suwaetal2015}
and \citet{Muelleretal2018} for ultrastripped SN~Ic
progenitors seems to support these theoretically expected relations.

In this paper we present a large sample of 45 explosion calculations in 2D and 3D
for neutrino-powered electron-capture SNe (ECSNe) of an O-Ne-Mg-core progenitor.
Our results reflect the described trends much more prominently.
Because of the extremely tenuous H/He envelope around 
the degenerate core, none of our calculations produced a NS kick of more than a few
km\,s$^{-1}$. This result is in line with considerations by \citet{Podsiadlowskietal2004},
who hypothesized that ECSNe give birth to NSs with low kick velocities.
However, detailed analyses of hydrodynamic explosion models
\citep{Schecketal2004,Schecketal2006,Wongwathanaratetal2010,Wongwathanaratetal2013,Nordhausetal2010}
revealed a different mechanism than that argued by \citet{Podsiadlowskietal2004}.
\citet{Schecketal2004} found that NS kicks are caused by asymmetries in the 
mass ejection of SN explosions rather than anisotropies in the neutrino emission.
In ECSNe the low explosion energy and the extremely rapid outward expansion of SN shock
and postshock layer (all of which are linked to the steep density decline outside of
the degenerate core) are detrimental to the possibility of large NS kicks.\footnote{
According to our present understanding of the neutrino-driven SN 
mechanism and NS kicks, it is not crucial for the kick magnitude whether the 
explosion sets in 
on a short post-bounce timescale (``promptly'') or with some longer delay.
Also low-mass stars with small iron cores and tenuous surrounding shells can exhibit
long delays before the explosion sets in, whereas more massive progenitors with 
larger iron cores and dense surrounding shells can explode more rapidly due to 
high neutrino luminosities caused by high mass accretion rates onto the nascent NSs 
\citep[see, e.g.,][]{Summaetal2016,Ertletal2016,Sukhboldetal2016,Couch2017,
Vartanyanetal2018,Ottetal2018}. Moreover,
large explosion asymmetries and large NS kicks do not
require a particularly long delay of the onset of the explosion, because
low-order (dipolar and quadrupolar) asymmetries due to the SASI or convection
can grow on similarly short timescales of only $\sim$100\,ms. Instead, the 
growth rates of both instabilities depend strongly on the time evolution
of the shock radius, which determines the conditions in the postshock
accretion flow. A larger shock radius leads to a slow accretion flow and thus
favors the growth of convection, a small shock radius and correspondingly fast 
accretion flow provide favorable conditions for the growth of the SASI 
\citep[see, e.g.,][]{Foglizzoetal2006,Foglizzoetal2007,Schecketal2008,Burrowsetal2012,
Fernandezetal2014,Fernandez2015,Foglizzoetal2015,Janka2017a}.}

Our results have important implications for the discussion of the Crab SN and
its stellar origin, for which an O-Ne-Mg core progenitor has been proposed
\citep{Nomotoetal1982,Hillebrandt1982}. Such a connection seems to be compatible with the
He-rich chemical composition of the nebula and a small amount of ejected
iron-group material \citep{Nomotoetal1982,MacAlpineSatterfield2008,Wanajoetal2011,Smith2013}, 
a small ejecta mass \citep{DavidsonFesen1985,Fesenetal1997}, low kinetic energy
\citep{YangChevalier2015,Kitauraetal2006,Dessartetal2006,Fischeretal2010,Radiceetal2017},
and the characteristics of the light curve \citep{Tominagaetal2013,Smith2013}. 
However, the small NS kick velocities of only a few km\,s$^{-1}$ obtained for ECSN 
explosions in our work are in conflict with the $\sim$160\,km\,s$^{-1}$ deduced from
observations for the spatial velocity of the Crab pulsar 
\citep{Hester2008, Kaplanetal2008}. 
This disaccord points to several possible consequences: If the Crab 
Nebula was born in an ECSN, the motion of the Crab pulsar either requires
a non-hydrodynamic acceleration mechanism, for example anisotropic neutrino emission
or the post-natal electromagnetic rocket effect \citep{HarrisonTademaru1975}, 
or the pulsar's motion is attributed to the disruption of a close binary system during
the SN explosion \citep[Blaauw effect;][]{Blaauw1961}. In the latter case, however, the 
structure and internal dynamics of the Crab Nebula with its pulsar may have to be
revised. Alternatively, the mismatch between our calculations for ECSNe and the 
observed space velocity of the Crab pulsar might imply that Crab is the relic of
the explosion of a low-mass iron-core progenitor.
Since such progenitors can possess considerably higher values of
the core compactness (i.e., more shallow density profiles around the degenerate core) 
than O-Ne-Mg-core progenitors, the hydrodynamical kick mechanism might account for
the proper motion of the Crab pulsar in this case. We will critically assess these
possibilities in a detailed discussion.

In Sect.~\ref{sec:num} we summarize our numerical methods and input physics,
in Sect.~\ref{sec:models} we describe our set of computed models and explosion results,
in Sect.~\ref{sec:kicks} we present our results for the corresponding NS kicks, in 
Sect.~\ref{sec:discussion} we discuss the implications for Crab, and
Sect.~\ref{sec:conclusions} contains a summary and our conclusions.

\section{Computational approach}
\label{sec:num}
\subsection{Numerical code and input physics}
Our numerical code, termed \textsc{Prometheus-HotB}, is a derivative of the version 
described in more detail by \citet{Wongwathanaratetal2013} and upgraded by improvements 
for the equation of state (EoS) and for the treatment of nuclear burning as
explained in \citet{Ertletal2016}.
\textsc{Prometheus-HotB} contains the hydrodynamics module \textsc{Prometheus}, 
which is based on an explicit, higher-order finite-volume Riemann solver for 
the Eulerian hydrodynamics equations \citep{Fryxelletal1989}.
Self-gravity is accounted for by using a multipole expansion of Poisson's 
equation in its integral form \citep{MuellerSteinmetz1995}.
For our simulations, we restrict ourselves to the monopole moment of the 
gravitational potential, which includes a correction for general relativistic
gravity effects according to \citet{Mareketal2006} and \citet{Arconesetal2007}. 
This restriction to
the monopole term is no constraint for our calculations of nonrotating models,
because in this case the gravitational potential is largely dominated by the
spherically symmetric NS of $\sim$1.35\,$M_\odot$, while only minuscule 
nonspherical devitations can be expected by density inhomogeneities in the
tiny mass (at most $\sim$$2\times 10^{-2}$\,$M_\odot$) between NS surface 
and SN shock front. 

A grey ``ray-by-ray'' approximation is used for the neutrino transport as
implemented by \citet{Schecketal2006} in order to describe neutrino production and 
interactions in the neutrino-decoupling layers exterior to the highly 
neutrino-opaque core of the NS, which is replaced by an inner grid boundary
in our simulations (see Sect.~\ref{sec:grid}).
At densities above $\rho = 10^{11}\,\mathrm{g\,cm}^{-3}$, we employ the 
high-density equation of state (EoS) of \citet{LattimerSwesty1991} 
with a bulk incompressibility modulus of $K=220$\,MeV.
At lower densities, we apply an $e^{\pm}$, photon and baryon EoS from 
\citet{TimmesSwesty2000} for arbitrarily relativistic and degenerate 
leptons. Nuclear composition changes are self-consistently coupled
to the EoS and are described by nuclear statistical equilibrium for
16 nuclear species at temperatures $T\ge 5\times 10^9$\,K and computed with
a 14-nuclei $\alpha$-network (from \citealt{Mueller1986}
with reaction rates from the reaction library of \citealt{Thielemann1985})
for temperatures 
$9\times 10^8\,\mathrm{K}\le T < 5\times 10^9\,\mathrm{K}$. Below the lower 
temperature bound the nuclear composition is kept fixed. The burning 
network includes a tracer nucleus to represent neutron-rich 
products formed at conditions with an electron fraction $Y_e\le 0.49$
\citep{Wongwathanaratetal2013,Ertletal2016}. Feedback by the energy generation in 
nuclear reactions is fully taken into account in our hydrodynamics solver.

%
\begin{table}
\caption{Parameters values for the five model sets with chosen SN energies}
\setlength{\tabcolsep}{5pt}
\centering
{
\setlength{\extrarowheight}{3pt}
\begin{tabular}{lcccc}
\hline
\hline
\multirow{2}{2cm}{Model set} & $E_\mathrm{exp}$          &$\Delta E_{\nu,\mathrm{core}}^\mathrm{tot}$ & $R_\mathrm{ib}^\mathrm{f}$& $t_\mathrm{ib}$\\
                             & $[10^{49}\,\mathrm{erg}]$ &$[10^{51}\,\mathrm{erg}]$                   & $[\mathrm{km}]$           & $[\mathrm{s}]$\\
\hline
O3      & $\sim$3            & 0.2                           & 20                    & 2.0 \\
O5      & $\sim$5            & 3.6                           & 15                    & 1.0 \\
O8      & $\sim$8            & 17.9                          & 15                    & 0.7 \\
O12     & $\sim$12           & 35.8                          & 15                    & 0.5 \\
O16     & $\sim$16           & 53.6                          & 17                    & 0.3 \\

\hline
\end{tabular}}
\label{tab:settings}
\end{table}

\begin{table*}
\caption{Naming convention for 2D models (see Sect.~\ref{sec:com})}
\centering
{
\setlength{\extrarowheight}{3pt}
\begin{tabular}{cccccc}
\hline
\hline
O          & $\{3,5,8,12,16\}$         & -- & \{v,d\}         & \{h,i,l\}        & \#\\
\hline
O-Ne-Mg-   & $E_\mathrm{exp}$          &    & perturbed       & amplitude        & number \\
core       & $[10^{49}\,\mathrm{erg}]$ &    & quantity        & h $\equiv 1\%$   & of model\\
progenitor &                           &    & v $\equiv v_r$  & i $\equiv 0.5\%$ &\\
           &                           &    & d $\equiv \rho$ & l $\equiv 0.1\%$ &\\
\hline
\end{tabular}}
\label{tab:models}
\end{table*}

\subsection{Numerical grid and boundary conditions}
\label{sec:grid}
Our 2D simulations are performed with a polar coordinate grid, whereas for
the 3D simulations we
employ an axis-free Yin-Yang grid \citep{KageyamaSato2004}, implemented into
\textsc{Prometheus-HotB} by \citet{Wongwathanaratetal2010a} as an overlay of 
equatorial patches of two spherical polar grids, tilted by 90 degrees
relative to each other. The Yin-Yang grid avoids the coordinate singularity
along the polar axis of the spherical grid and thus permits computations with
considerably higher efficiency by relaxing the Courant-Friedrichs-Lewy (CFL)
condition on the time step.

We use 1310 zones in the radial direction for both 2D and 3D models.
For all simulations we employ the same non-equidistant spacing of the radial
grid points, similar to the grid described by \citet{Jankaetal2008}, 
in order to resolve the steep density gradient at the surface of the O-Ne-Mg core.
In the angular directions the grid is chosen to be equidistant both in 2D and 3D.
We apply an angular resolution of $1^{\circ}$ for the 2D simulations and of 
$3^{\circ}$ for our 3D simulations. This corresponds to 180 azimuthal zones in 2D
and $30(\theta)\times 90(\phi)$ angular zones on each of the two Yin-Yang patches
of the 3D grid. 

In both the 2D and 3D calculations we exclude the high-density, 
neutrino-opaque core of the newly formed NS and replace it by a
gravitating point mass at the grid center as well as an inner,
moving grid boundary at a prescribed, time-dependent radius, 
following the numerical treatment used already in previous publications
\citep{JankaMueller1996,Kifonidisetal2003,Schecketal2006,Arconesetal2007}. 
Thus excising the central region of the proto-NS (PNS) from the 
hydrodynamical domain relaxes the time-step constraint. Therefore
it permits us to compute a large set of 2D and 3D models for one second of
physical evolution time with reasonable computational resources. 
A full treatment from first principles would have been
computationally prohibitive, in particular in 3D, for gathering model
statistics over a larger sample of simulation runs. Moreover, the grid
boundary provides us with a simple handle to regulate the SN explosion energy
to preferred numbers by choosing suitable values of the free parameters involved
in our description of the boundary condition. 

Such a (relatively) free choice of the explosion
energy\footnote{With a chosen set of parameter values for
regulating the core-boundary luminosities the explosion energy exhibits some
variation because of the stochastic behavior of convectively buoyant 
neutrino-heated Rayleigh-Taylor plumes and their interaction with accretion 
downflows, in particular in 2D models. This feeds back into the neutrino
luminosities computed by the grey transport approximation that we apply 
outside of the excised central PNS core. Especially in the cases with the
lowest explosion energies the corresponding variations can reach up to 
about 20\% of the average value (see Table~\ref{tab:kick}).}
in our study is desirable not only because it allows us to account for
uncertainties of the ECSN energy on the observational side 
\citep[see, e.g.,][]{Tominagaetal2013,YangChevalier2015}
but also for remaining uncertainties in
the first-principle modeling of ECSN explosions. Theoretical uncertainties
are, for example, connected to
the still incompletely determined properties of the high-density EoS in hot
NSs and the corresponding neutrino opacities. Both have an influence on the 
exact energetics of the neutrino-driven wind of the PNS and, hence, also on
the energy of the neutrino-powered SN blast \citep{vonGroote2014}.
For our purpose, the use of a parametrized model to avoid
a full numerical treatment of the high-density NS core is therefore 
preferable, because we are interested in a large, representative 
study of the multi-dimensional dynamics of the SN ejecta for defined explosion
energies. Quantities that are directly influenced by the physics within
the nascent NS, such as the exact properties of the emitted neutrino signal
or the internal evolution of the NS itself, are not of interest for our 
study.

Following \citet{Schecketal2006} and \citet{Arconesetal2007} and motivated by
the shrinking of the NS core as it cools and deleptonizes by neutrino emission, 
we apply a contraction of the inner-boundary radius according to
\begin{equation}
R_{\rm{ib}}(t) = R_{\mathrm{ib}}^{\mathrm{f}} +
\left(R_{\mathrm{ib}}^{\mathrm{i}}-R_{\mathrm{ib}}^{\mathrm{f}}\right) 
\exp{(-t/t_{\mathrm{ib}})} \,.
\label{eq:rib}
\end{equation}
This prescription causes the inner grid boundary to retreat from an initial 
radius $R_{\mathrm{ib}}^{\mathrm{i}}$ to a final radius 
$R_{\mathrm{ib}}^{\rm{f}}$ with a characteristic exponential timescale 
$t_{\mathrm{ib}}$. 
For the progenitor studied in this work, we choose $R_{\mathrm{ib}}^{\rm{i}}=43.5$\,km.
This corresponds to an enclosed baryonic mass of 
$M_{\mathrm{encl}}^{\mathrm{b}} = 1.02$\,$M_\odot$ in the post-bounce 
initial state from which we start our multi-dimensional simulations. This enclosed
baryonic mass is kept constant with time, for which reason $R_\mathrm{ib}(t)$
tracks a fixed mass coordinate. The
hydrodynamic quantities that must be defined in the ghost cells at the inner 
boundary are chosen such that the condition of hydrostatic equilibrium is fulfilled
\citep[see][]{JankaMueller1996}.

At the outer boundary of the computational mesh at a radius of 
$R_{\mathrm{ob}} = 4.6\times 10^5$\,km, a free 
in/outflow boundary condition is applied.
This radius is picked such that the SN shock does not leave the computational
domain within the simulated evolution.

As in \citet{Schecketal2006} and \citet{Arconesetal2007}, we impose spherically 
symmetric luminosities for all neutrino species at the inner grid 
boundary.\footnote{We emphasize that spherically symmetric
boundary luminosities are imposed at a mass coordinate of 1.1\,$M_\odot$, i.e.
well inside of the PNS, which has a baryonic mass of at least 1.35\,$M_\odot$.
Our treatment of the neutrino transport through the outer 0.25\,$M_\odot$ of the 
PNS, which partially contain a convective layer that forms inside of the NS, 
can create neutrino emission and heating
asymmetries similar to other multi-dimensional SN simulations with neutrino 
transport. In addition, asymmetric accretion can add to neutrino-emission 
asymmetries. All of these neutrino asymmetries, however, are not responsible
for the growth of the hydrodynamic instabilities that lead to the asymmetric
SN explosions. This has been demonstrated in many previous works, e.g., by 
\citet{JankaMueller1996,Mezzacappaetal1998,MurphyBurrows2008,Nordhausetal2010b,Hankeetal2012},
where spherically symmetric neutrino ``light bulbs'' were employed and 
categorically largely asymmetric explosions were obtained nevertheless.}
These core luminosities are
regulated by a parameter $\Delta E_{\nu,\mathrm{core}}^\mathrm{tot}$, which 
determines the total gravitational binding energy extracted by neutrino radiation
from the NS core. In the present study we use three parameters of the NS-core
model, namely $\Delta E_{\nu,\mathrm{core}}^\mathrm{tot}$, 
$R_{\mathrm{ib}}^{\mathrm{f}}$, and $t_{\mathrm{ib}}$ (Table~\ref{tab:settings})
to regulate the explosion energies of our computed ECSN models
(see Sect.~\ref{sec:models}). Although these parameters have the physical 
meaning mentioned above and discussed in detail by \citet{Schecketal2006},
their values are varied here beyond the ranges suggested by self-consistent
simulations of the NS-core evolution. We therefore discourage from a deeper
interpretation of the numbers given in Table~\ref{tab:settings}. They are
picked merely for the goal to produce desired values of the SN explosion energy
by neutrino heating. The corresponding combinations of parameter values are
not unambiguous, i.e., other choices can be equally suitable to obtain the 
same explosion energies.

\subsection{Seed perturbations}
\label{sec:seeds}
Because of the lack of multi-dimensional, self-consistently computed
stellar progenitor models, we have to apply artificial perturbations
to break spherical symmetry in the collapsing stellar matter and thus
to trigger the growth of nonradial hydrodynamic instabilities in unstable
regions. Following previous works
\citep[e.g.,][]{JankaMueller1996,Schecketal2006,Arconesetal2007,Wongwathanaratetal2013,Hankeetal2013},
both velocity and density profiles are considered for perturbation at
various amplitudes (see Table~\ref{tab:models}) to ensure that the evolving
stochasticity is relatively robust to the chosen form of underlying
perturbations. Indeed we do not observe any qualitative differences in the
outcomes caused by different perturbation patterns and amplitudes. While
for the 2D simulations the available computational resources permit us to test
a wider variety of seed perturbations, applied to either density or radial
velocity in random cell-to-cell variations with defined amplitudes,
the computational costs of 3D models allow us to employ merely one
perturbation setting for all 3D models.

\begin{figure}[!]
\centering
\includegraphics[width=1.\hsize]{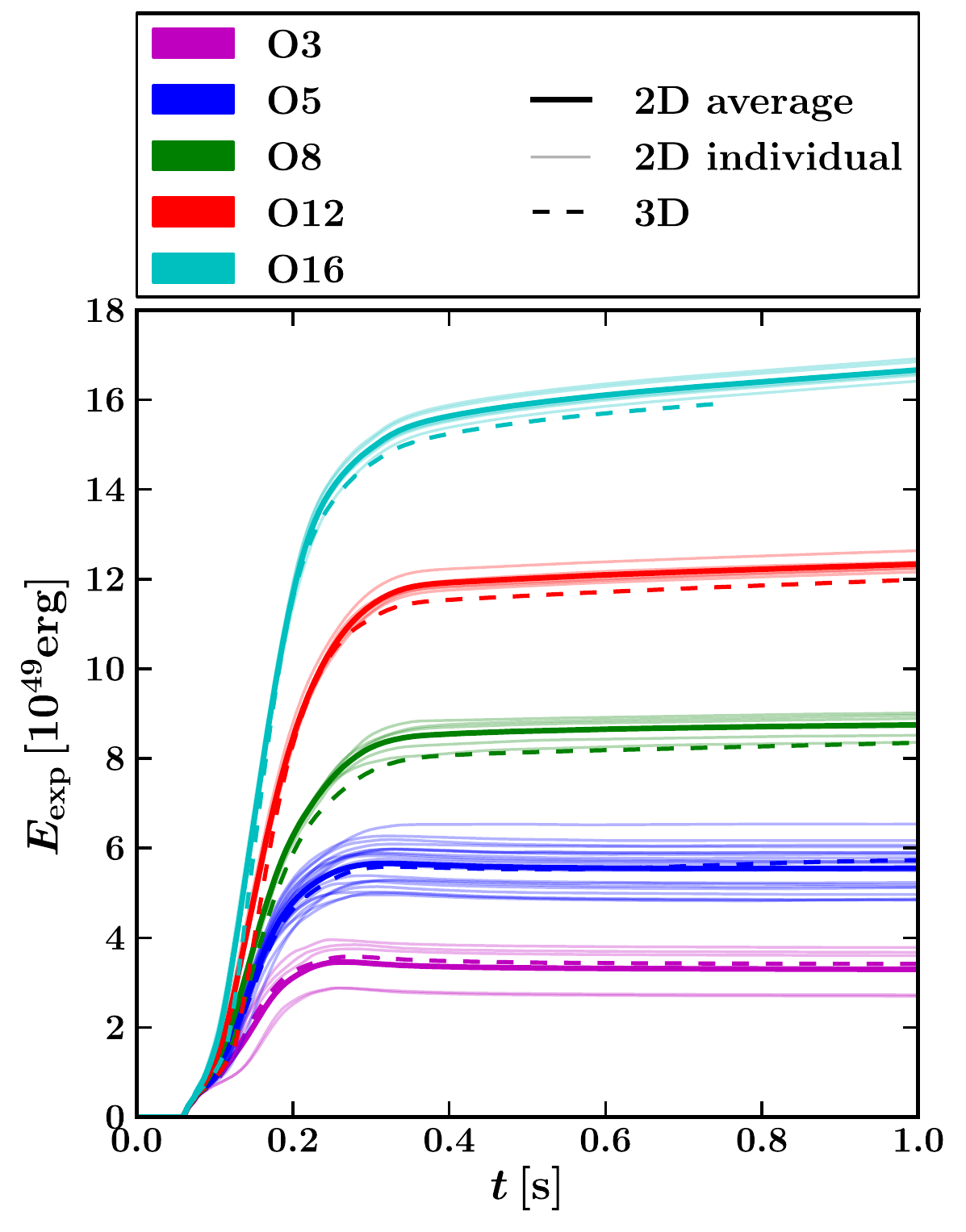}
\caption{Time evolution of the (diagnostic) explosion energies for 2D models
(thin solid lines), their average (thick solid lines), and 3D models (dashed
lines). Each color represents a fixed set of parameter values
(see Table~\ref{tab:settings}). At the end of the simulations, the
gravitational binding energy of overlying stellar material ahead of the
SN shock is negligible. Therefore the final diagnostic energy is essentially
equal to the explosion energy.}
\label{fig:eexp}
\end{figure}

\begin{figure*}[!]
\centering
\includegraphics[width=0.49\textwidth]{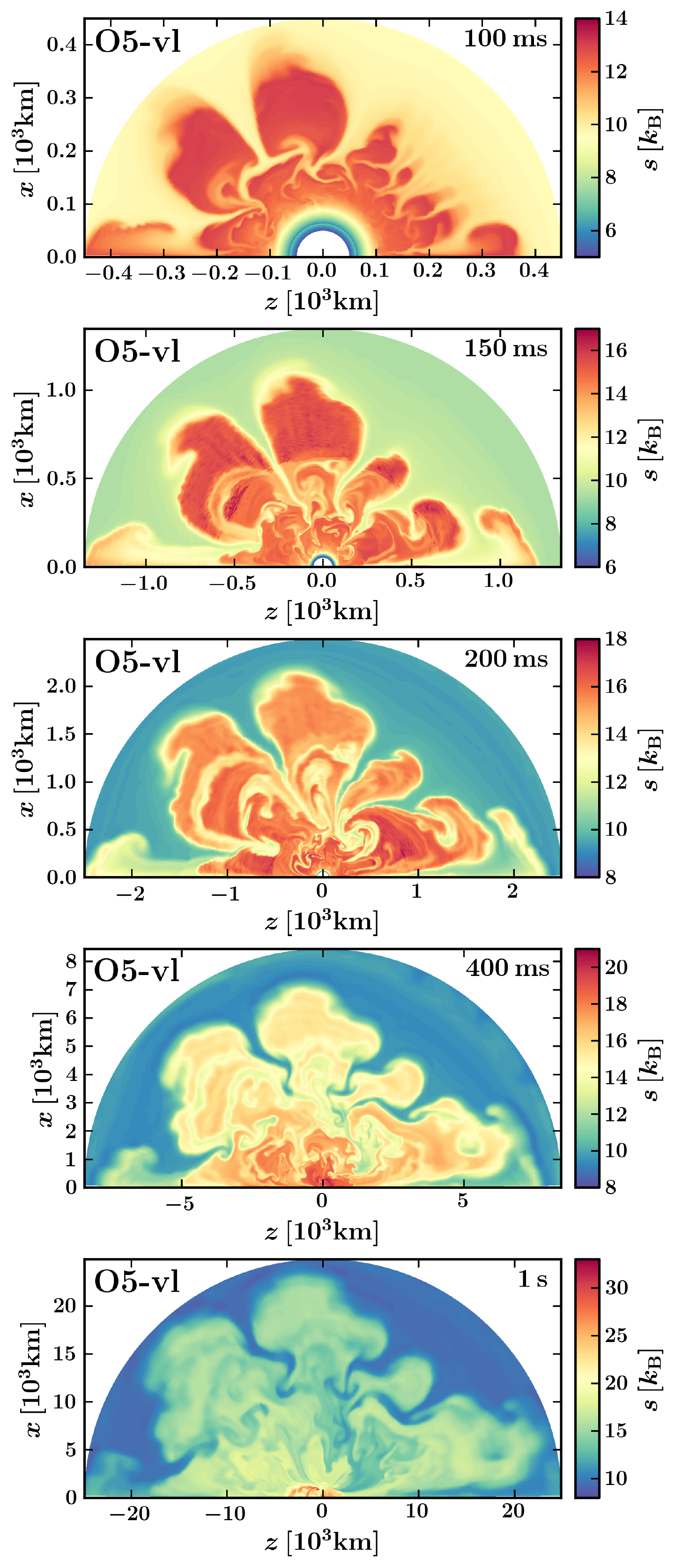}
\includegraphics[width=0.49\textwidth]{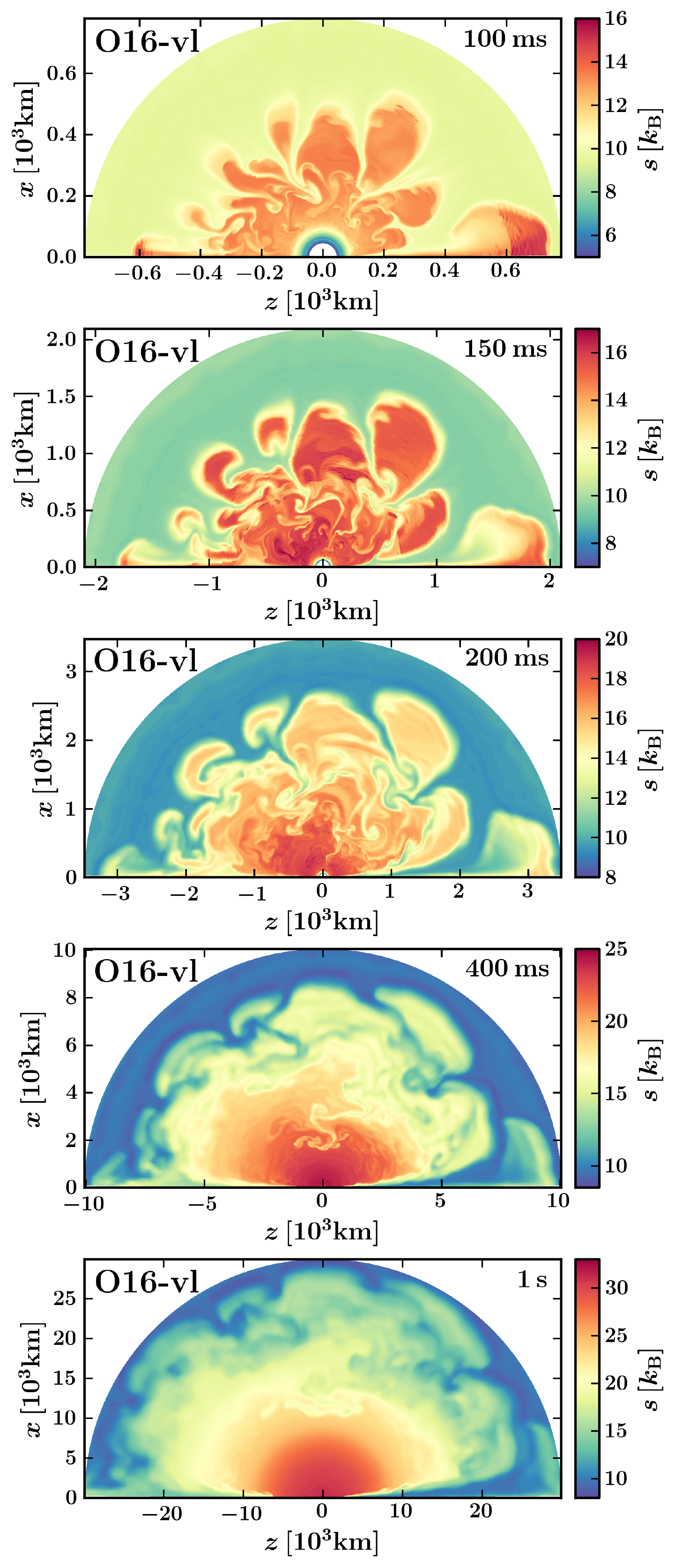}
\caption{Distributions of the entropy per nucleon 
at 100, 150, 200, 400\,ms and 1\,s
\textit{(from top to bottom)} for 2D models O5-vl
\textit{(left column)} and O16-vl \textit{(right column)}. By the time
convective, high-entropy plumes start rising 
on a scale of a few hundreds of kilometers
at $\sim$100\,ms, the SN shock has passed a radius of $10^4\,\mathrm{km}$
already. In the more energetic model O16-vl, the
high-entropy plumes grow faster than in the lower-energy model O5-vl.
Because of the higher neutrino luminosity imposed at the inner grid
boundary, model O16-vl quickly develops a spherically symmetric
neutrino-driven wind, except for distinct downdrafts at both poles.}
\label{fig:2dsto}
\end{figure*}

%
\section{Computed models}
\label{sec:models}
\subsection{Model sets}
\label{sec:com}
We investigate the explosion dynamics of an $8.8\,\msun$ star with an O-Ne-Mg 
core prior to collapse (\citealt{Nomotoetal1984, Nomotoetal1987}; 
plots of the progenitor structure, also in comparison to 
iron-core models and more modern O-Ne-Mg core progenitors, can be found in
\citealt{Jankaetal2008,Jankaetal2012,Jonesetal2013,Mueller2016,Janka2017a,CerdaDuran2018}).\footnote{
The detailed structure of the tenuous H/He envelope that surrounds the degenerate
core does not have any significant impact on the outcome of our simulations as 
long as there is a density gradient over several orders of magnitude between the 
degenerate core and the envelope. In \citet{Jankaetal2008} two different density 
profiles were used (see figure~1 there), which both led to the same explosion 
behavior in 1D and 2D simulations, despite the fact that one of the envelopes
had a density of $\gtrsim$100\,g\,cm$^{-3}$ at its base and a shallow density 
profile following about $r^{-3}$ outside, and thus locally up to 8 orders
of magnitude higher densities than in the other case.}
Rotation is not accounted for in our study.\footnote{ 
This is well justified because even for a rotation
period of $\sim$19\,ms as estimated for the Crab pulsar 
\citep{BejgerHaensel2003,ManchesterTaylor1977} the corresponding Rossby number
\citep[see, e.g.,][equation~7 there]{Summaetal2018},
i.e., the ratio of the radial expansion velocity of buoyant Rayleigh-Taylor 
plumes divided by the rotation velocity in the convective region (typically
at radii larger than $\sim$100\,km) is much larger than unity, for which reason
the growth of hydrodynamic instabilities is basically unperturbed by rotation.}

The collapse and early post-bounce evolution of the considered ECSN progenitor
were simulated in spherical 
symmetry by \citet{Huedepohletal2010} using the \textsc{Prometheus-Vertex} code 
\citep{RamppJanka2002}, which, unlike \textsc{Prometheus-HotB},
includes the whole core without parametrized
inner boundary condition and which contains all relevant microphysics
(in particular the electron-capture rates and a special treatment of highly
degenerate conditions) that are needed to follow the core-collapse and
shock-formation phases accurately.
At $18\,\mathrm{ms}$ post bounce, we adopt the hydrodynamic model state
from the \textsc{Prometheus-Vertex} calculation and track the subsequent
evolution by multi-dimensional simulations with the \textsc{Prometheus-HotB} code.
The mapping from one simulation to the other including a change of
the numerical grid (which has essentially the same radial resolution in both cases)
and of the neutrino treatment does not produce any noticeable numerical artifacts.
This allows us to perform 40 explosion runs in 2D and five more models in 3D 
until 1\,s physical time. Note that in the whole of our paper, unless stated
otherwise, time is always defined by the \textit{simulated time of evolution}
after taking over the computation with \textsc{Prometheus-HotB}.
For the true post-bounce time 18\,ms have to be added to this clock. The large
number of simulations enables us to explore the statistical fluctuations of the
NS kicks.

The free parameters that describe the core-neutrino source in our setup,
$R_{\mathrm{ib}}^{\mathrm{f}}$, $t_{\mathrm{ib}}$, and 
$\Delta E_{\nu,\mathrm{core}}^\mathrm{tot}$ (see Sect.~\ref{sec:grid}), 
are calibrated such that our models reproduce basic explosion features of
ECSNe found in fully self-consistent simulations 
\citep{Kitauraetal2006,Jankaetal2008,Huedepohletal2010,Fischeretal2010,
vonGroote2014,Radiceetal2017}, 
in particular the explosion energy and the time of the onset of the SN blast.
We choose five parameter triples for covering the possible range of explosion 
energies suggested by the mentioned full-scale ECSN models and compatible
with SN~1054 according to a detailed analysis of the Crab remnant 
\citep{YangChevalier2015}.
In Table~\ref{tab:settings} the corresponding values of the three relevant
calibration parameters are listed for our five model sets.

We use the following naming convention for the computed models in these
five sets (see also Table~\ref{tab:models}):
The initial letter ``O'' stands for ``O-Ne-Mg core'' and emphasizes that 
we consider the same progenitor star for all calculations.
The associated number from the set $\{3,5,8,12,16\}$ represents the approximate 
value of the explosion energy (in units of $10^{49}$\,erg). 
For the 2D models, the next two letters specify how the initial model 
is randomly perturbed (cell-by-cell) at the beginning of our simulation in 
order to break spherical symmetry and thus to initiate the growth of nonradial
hydrodynamic instabilities.
The first letter represents the perturbed quantity, i.e., either
the radial velocity $v_r$ (``v'') or the density $\rho$ (``d''). 
In the case of radial velocity
perturbations, conservation of the total (i.e., internal plus kinetic) energy
is enfored, in the case of density perturbations, conservation of the total
mass is ensured. The
amplitude applied for the initial perturbation is indicated by the second letter, 
``h'' (high), ``i'' (intermediate), and ``l'' (low) for an amplitude of 1\%, 
0.5\%, and 0.1\%, respectively. 
Finally, models computed for identical parameter settings are numbered 
sequentially. To facilitate orientation, we summarize our naming convention
in Table~\ref{tab:models}.
For all of the models computed in 3D, we apply an initial perturbation of
amplitude 0.1\% in the radial velocity. These models can be identified by the 
suffix ``3D'' in the model name. 
In our 2D simulations, both velocity and density profiles are
considered for perturbations with various amplitudes in order to verify that the
evolving stochasticity of the hydrodynamic instabilities is basically robust
to different types of perturbations. Indeed, we do not observe
any qualitative differences in the outcomes caused by different quantities and 
patterns of perturbation. For our 3D models we cannot provide the same variety
of cases as for the 2D models and employ merely one perturbation setting for 
all 3D simulations.
A ``set of models'' is defined by all explosion simulations that share the 
same values of the calibration parameters and, thus, develop very similar 
explosion energies, independent of their specific perturbation pattern.

\begin{figure*}[!]
\centering
\includegraphics[width=0.45\textwidth]{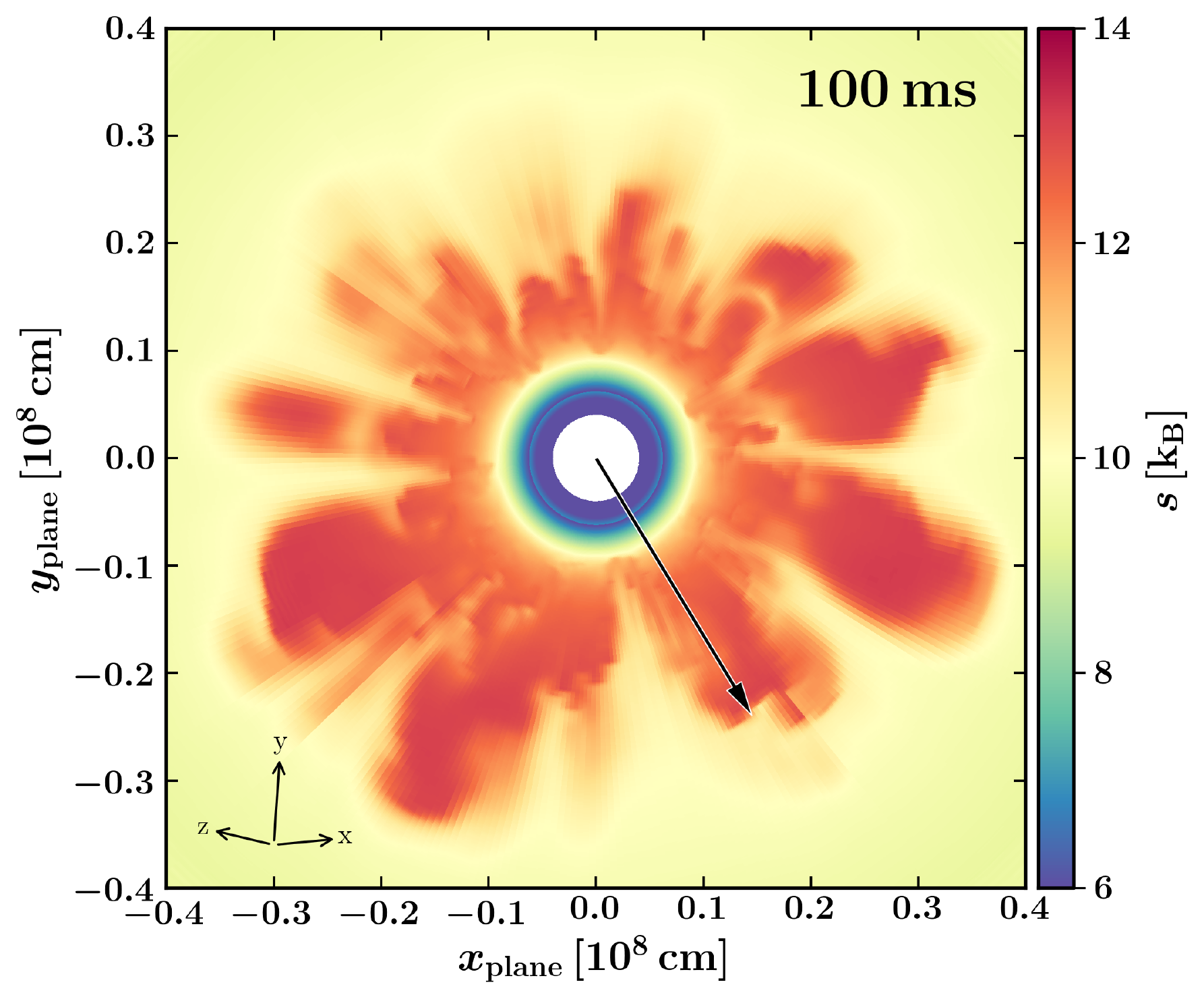}
\includegraphics[width=0.45\textwidth]{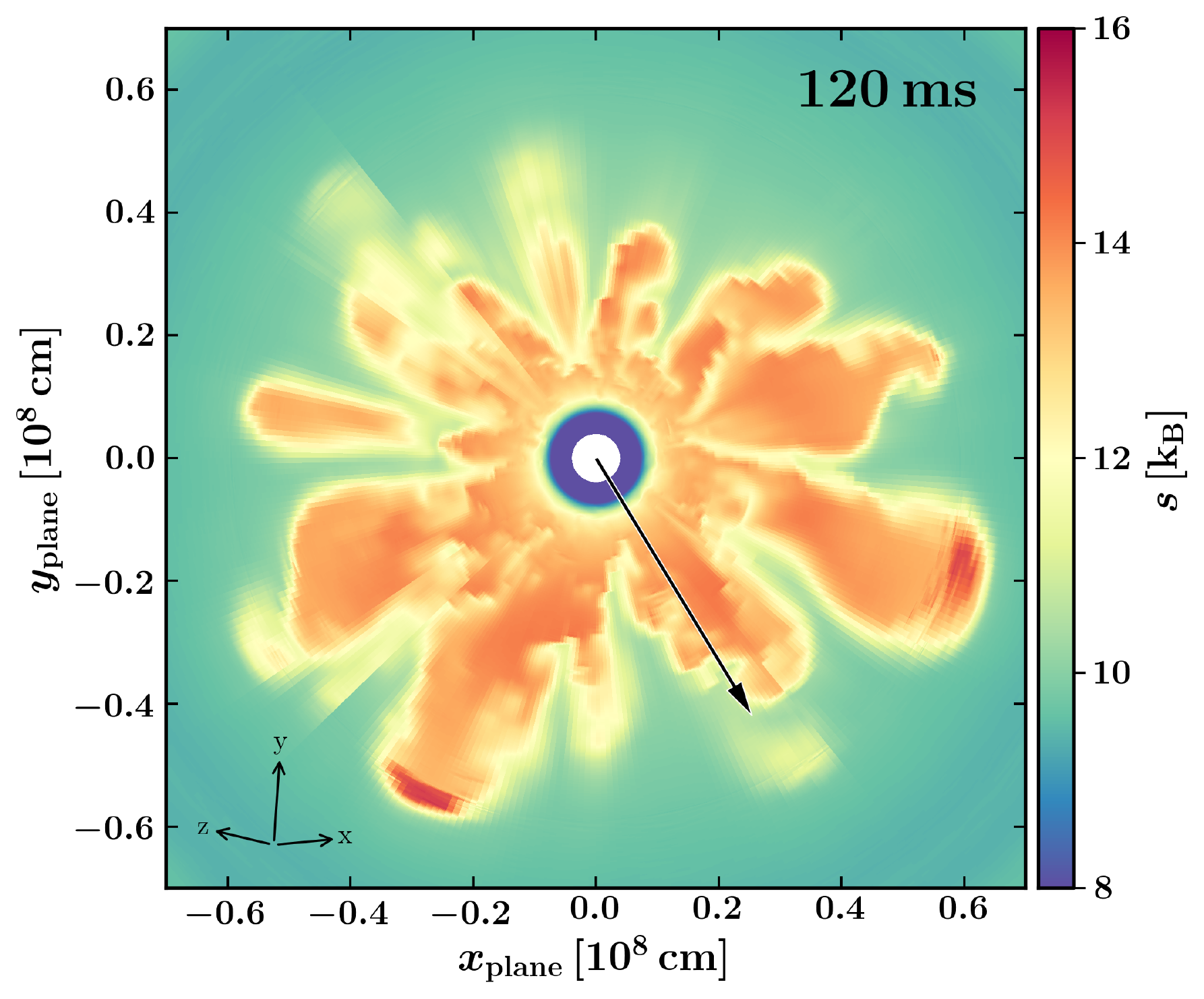}\\
\includegraphics[width=0.45\textwidth]{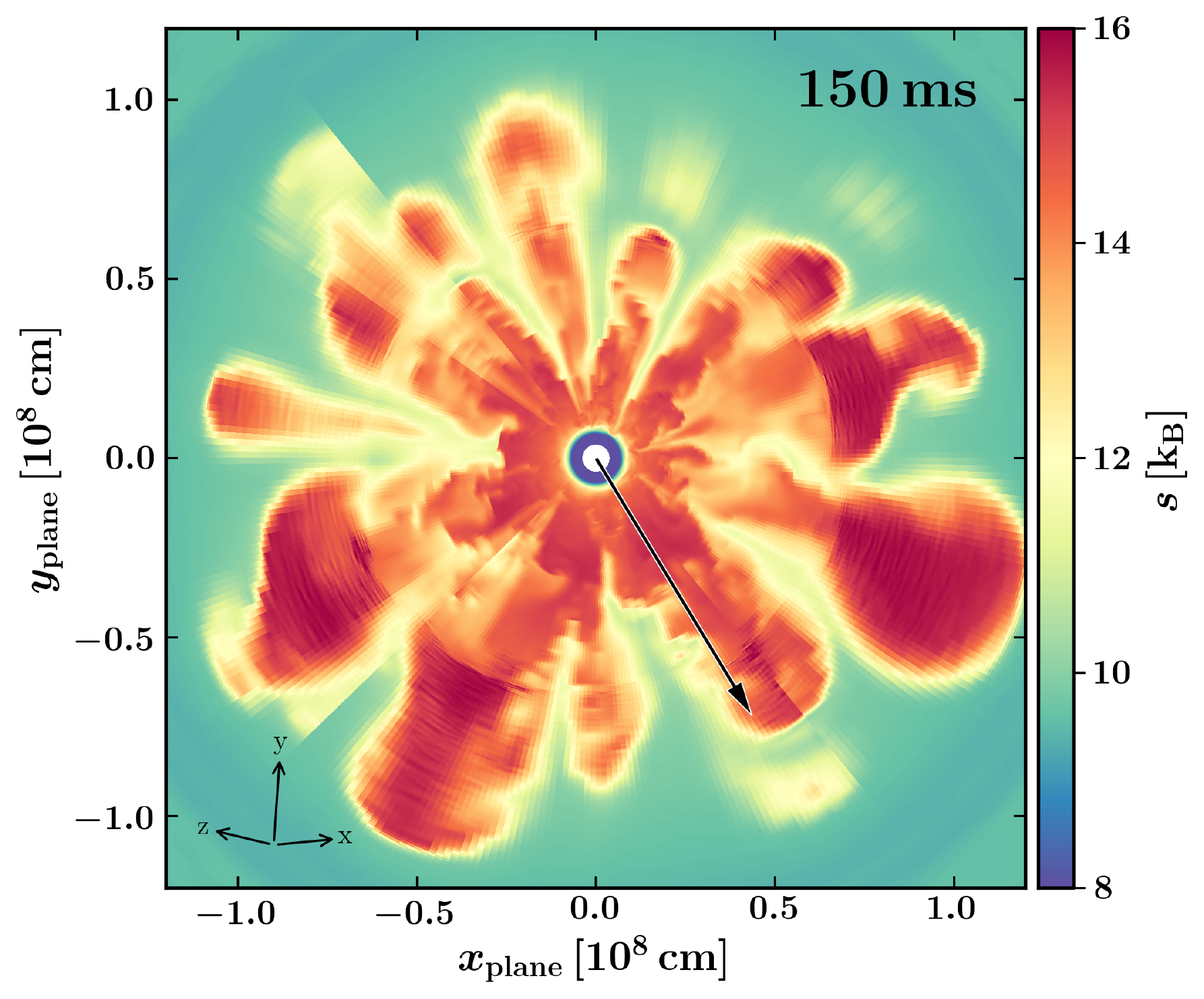}
\includegraphics[width=0.45\textwidth]{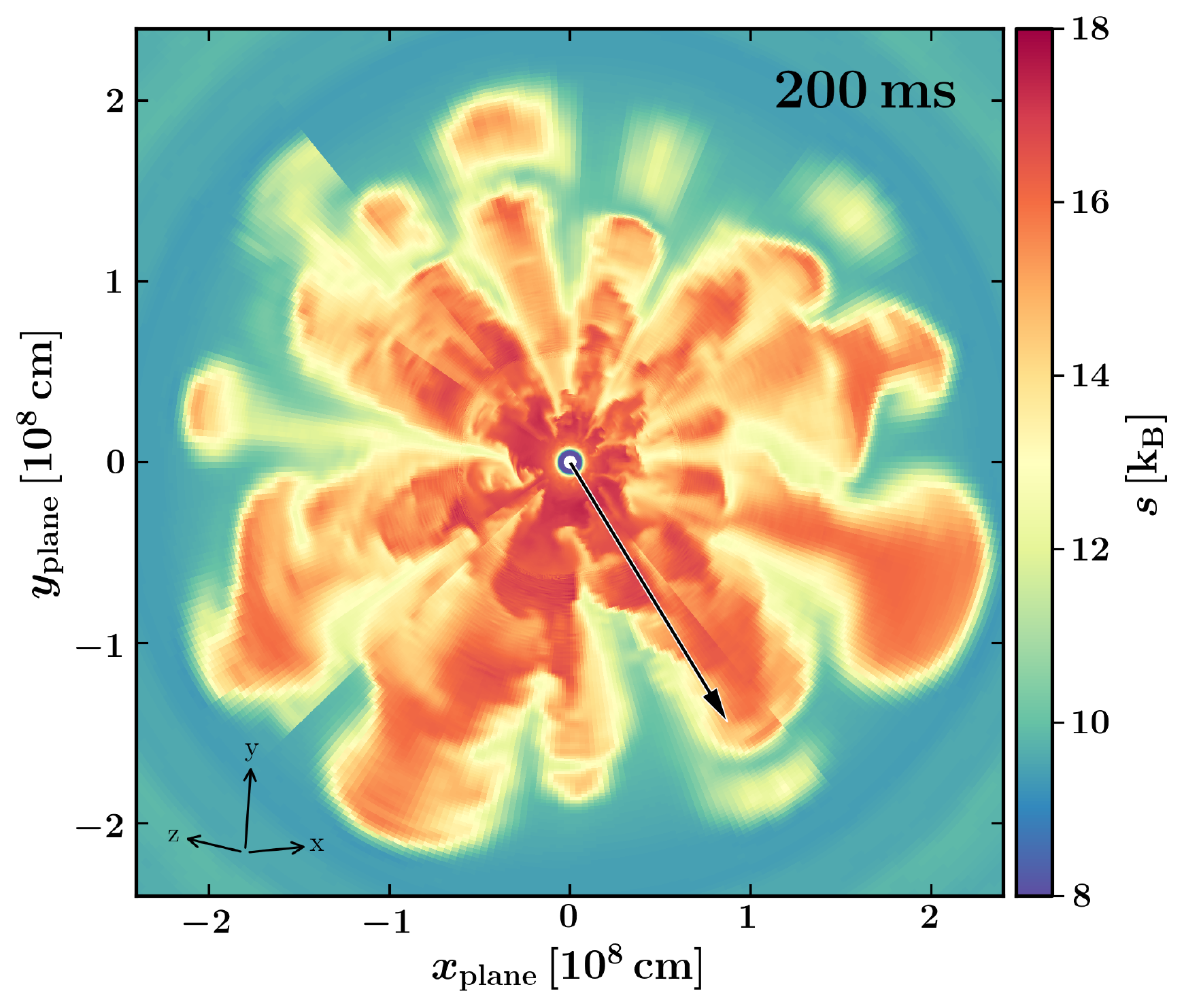}\\
\includegraphics[width=0.45\textwidth]{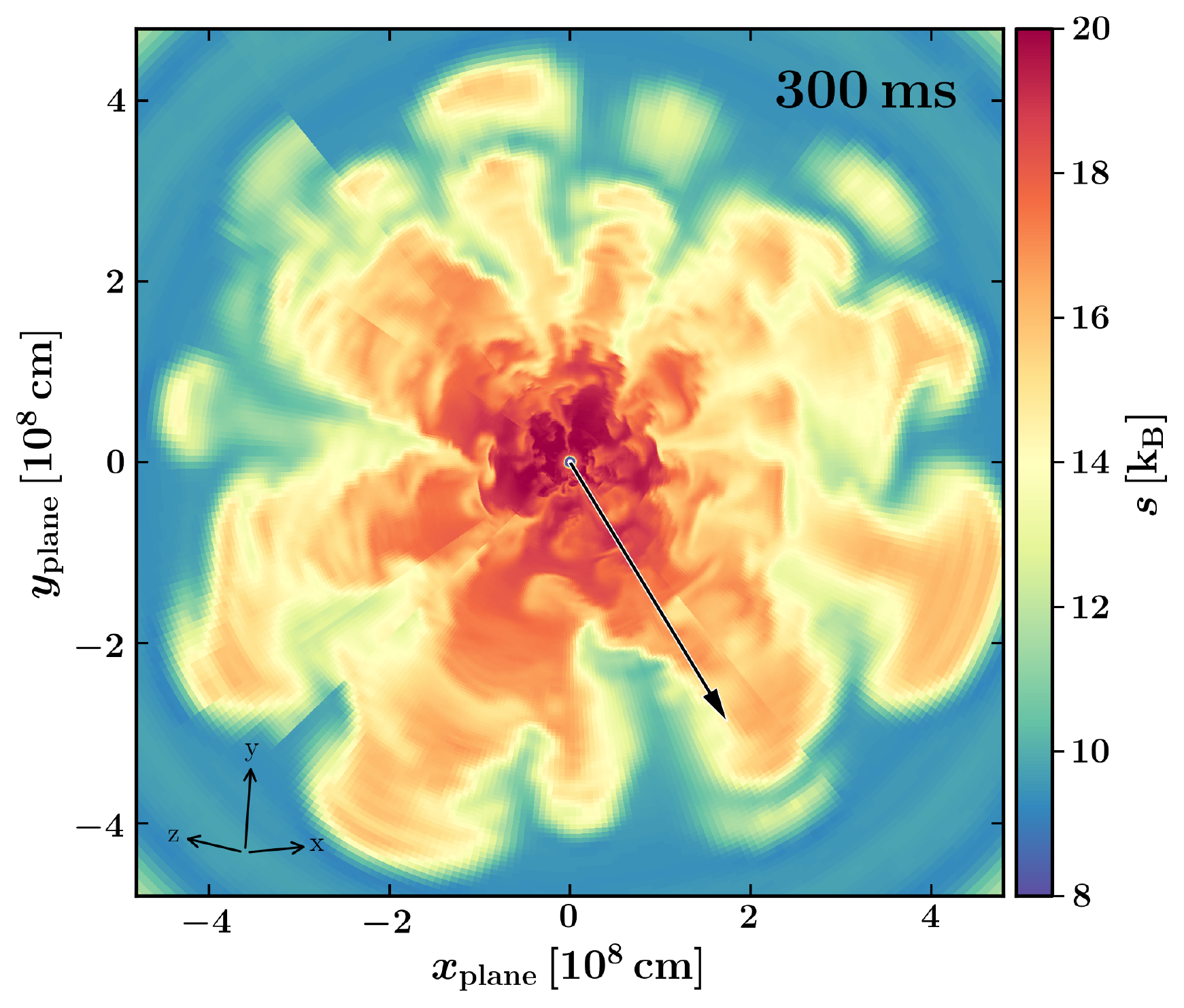}
\includegraphics[width=0.45\textwidth]{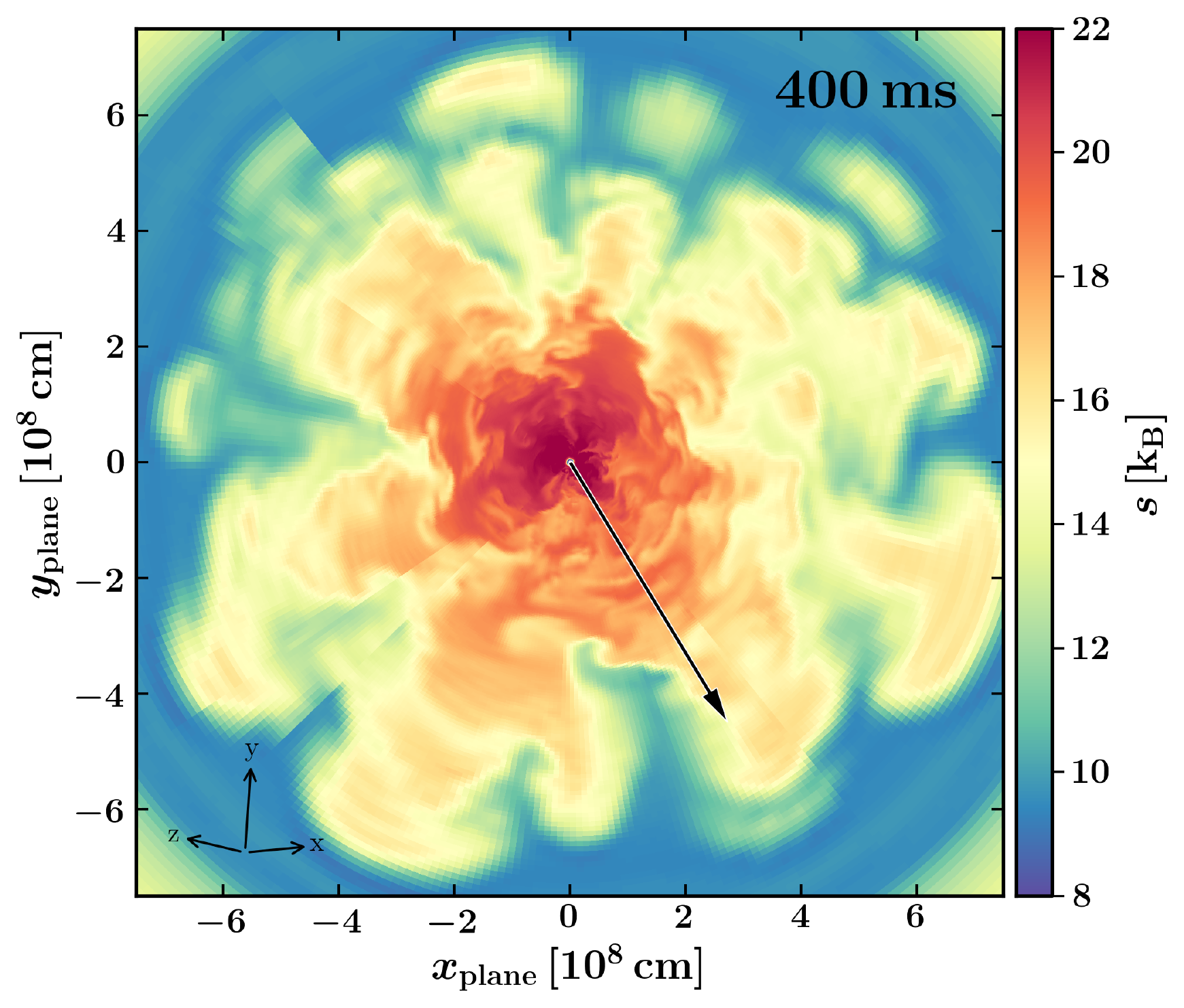}
\caption{Entropy per nucleon in slices displaying the early time evolution
of model O8-3D. The cross-sectional plane is spanned by a vector in the
(final) direction of the NS kick at 1\,s, indicated by the
black arrow, and the vector $(1,0,-1)$ in the coordinate system
represented by the tripod in the lower left corner of each panel.
\textit{Upper left panel:} After 100\,ms, buoyant plumes have grown to
radii of $\sim$300\,km. By this time, the shock has already advanced to a
distance of more than $10^4$\,km.
\textit{Following panels:} Within the first 400\,ms, the plumes swell to
radii of several 1000\,km; however, they stay far behind the rapidly expanding
shock. High-order multipoles dominate the pattern of the growing Rayleigh-Taylor
mushrooms, and a preferred direction for the NS kick cannot yet be clearly
identified from the ejecta asymmetry. Note that the interpolation of the
data from the Yin-Yang grid onto the chosen cut plane creates some plotting
artifacts, which do not occur in Fig.~\ref{fig:elena_plots}.}
\label{fig:3dsto}
\end{figure*}

\begin{figure*}[!]
\centering
\includegraphics[width=0.45\textwidth]{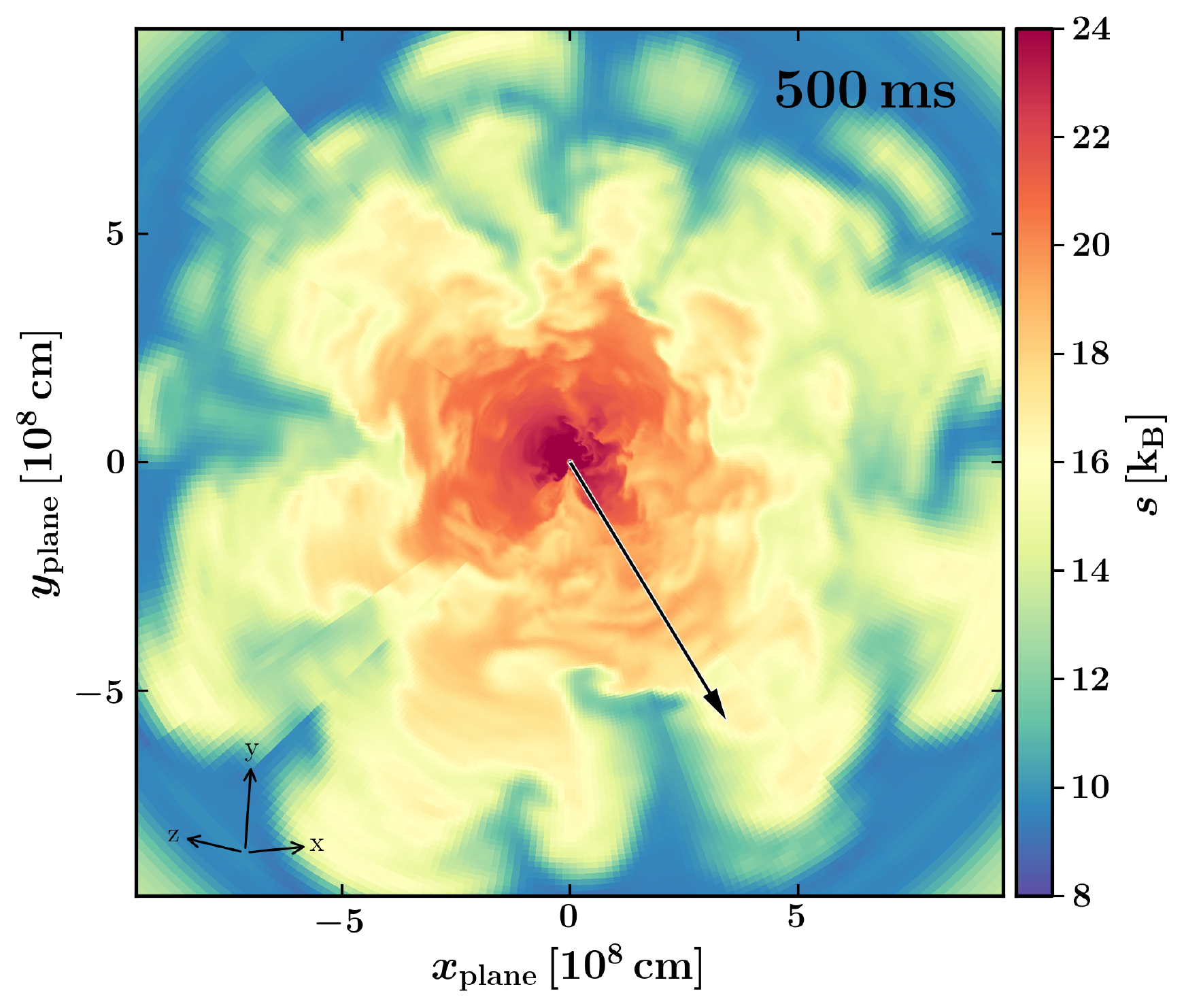}
\includegraphics[width=0.45\textwidth]{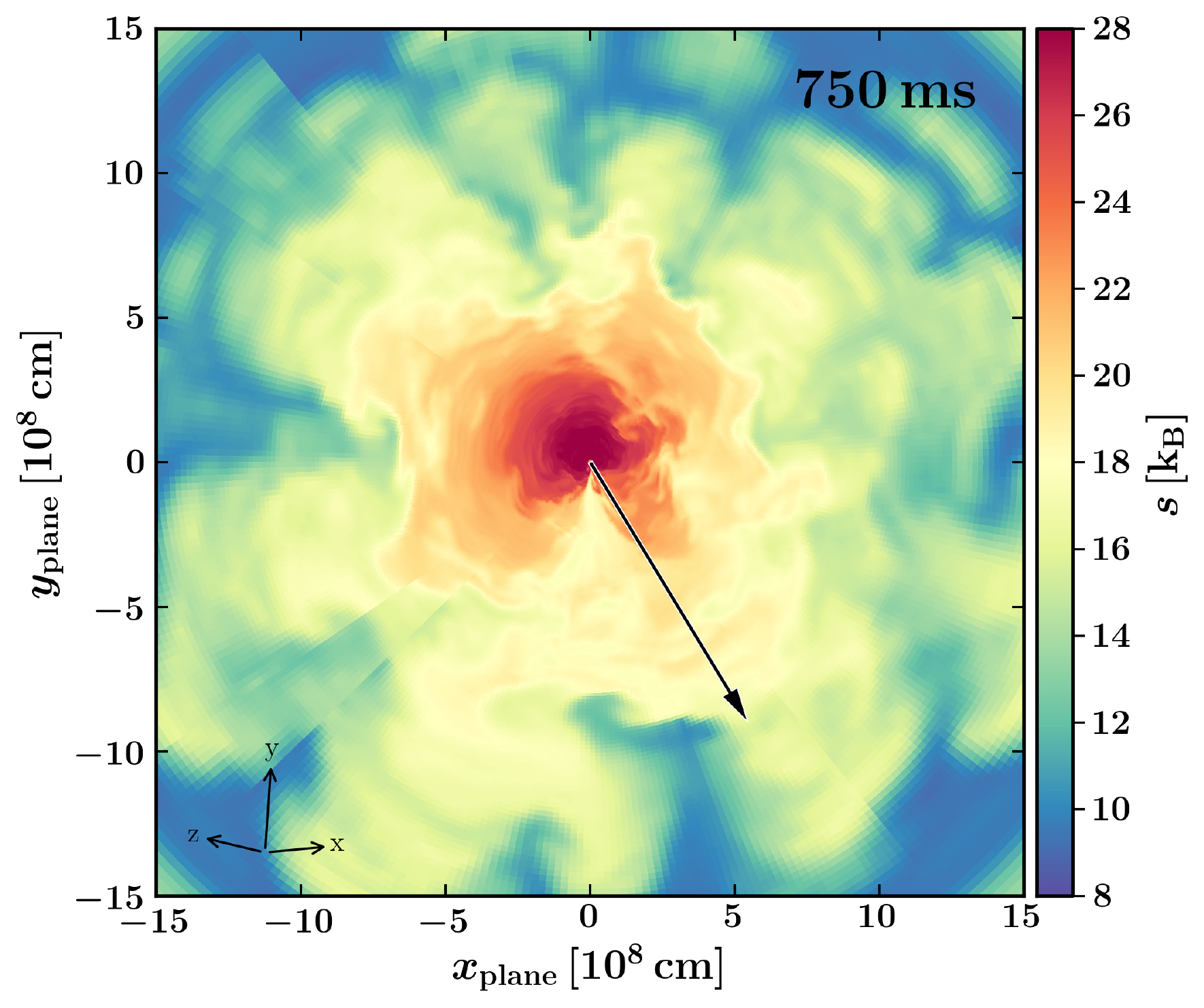}\\
\includegraphics[width=0.45\textwidth]{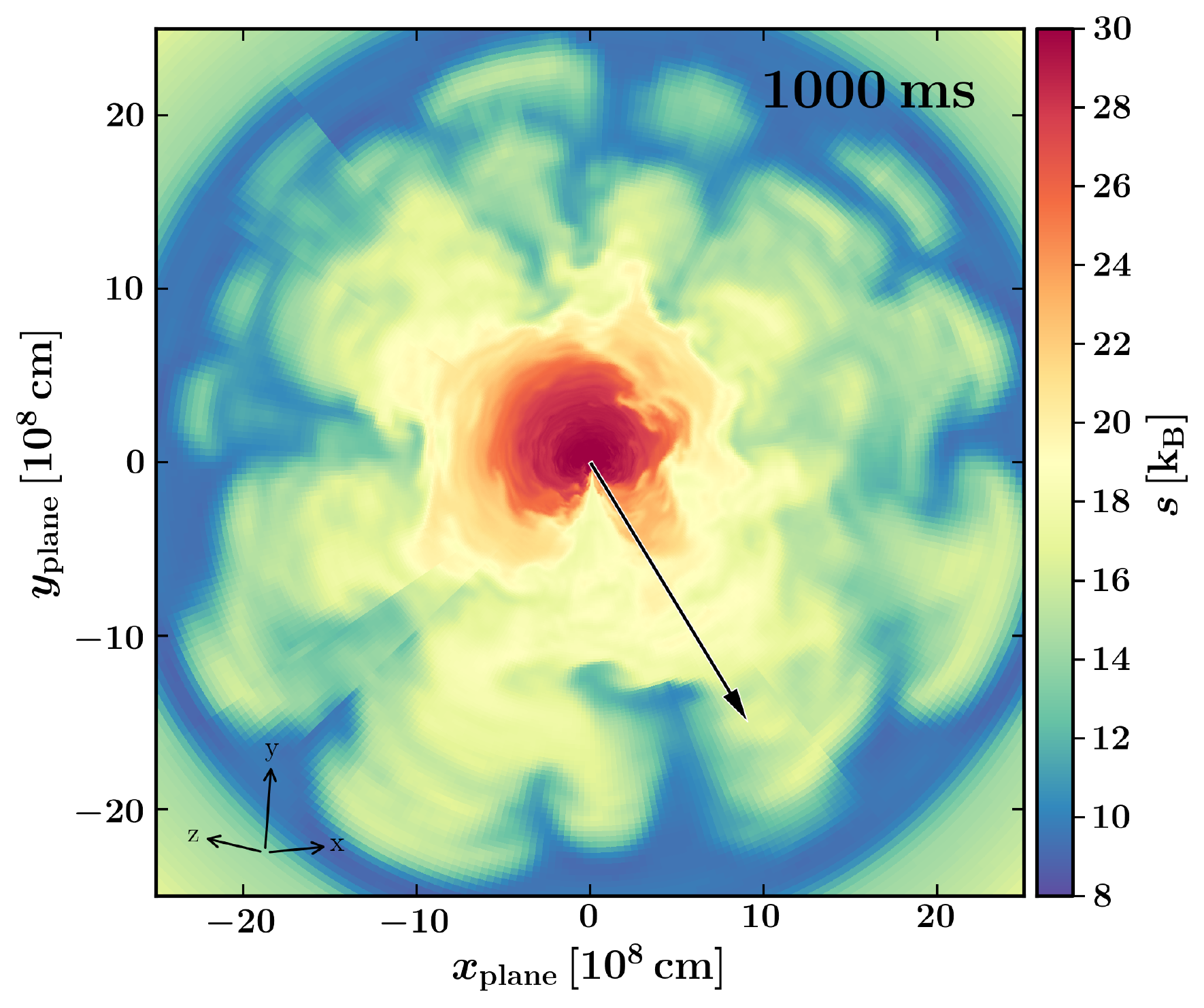}
\includegraphics[width=0.45\textwidth]{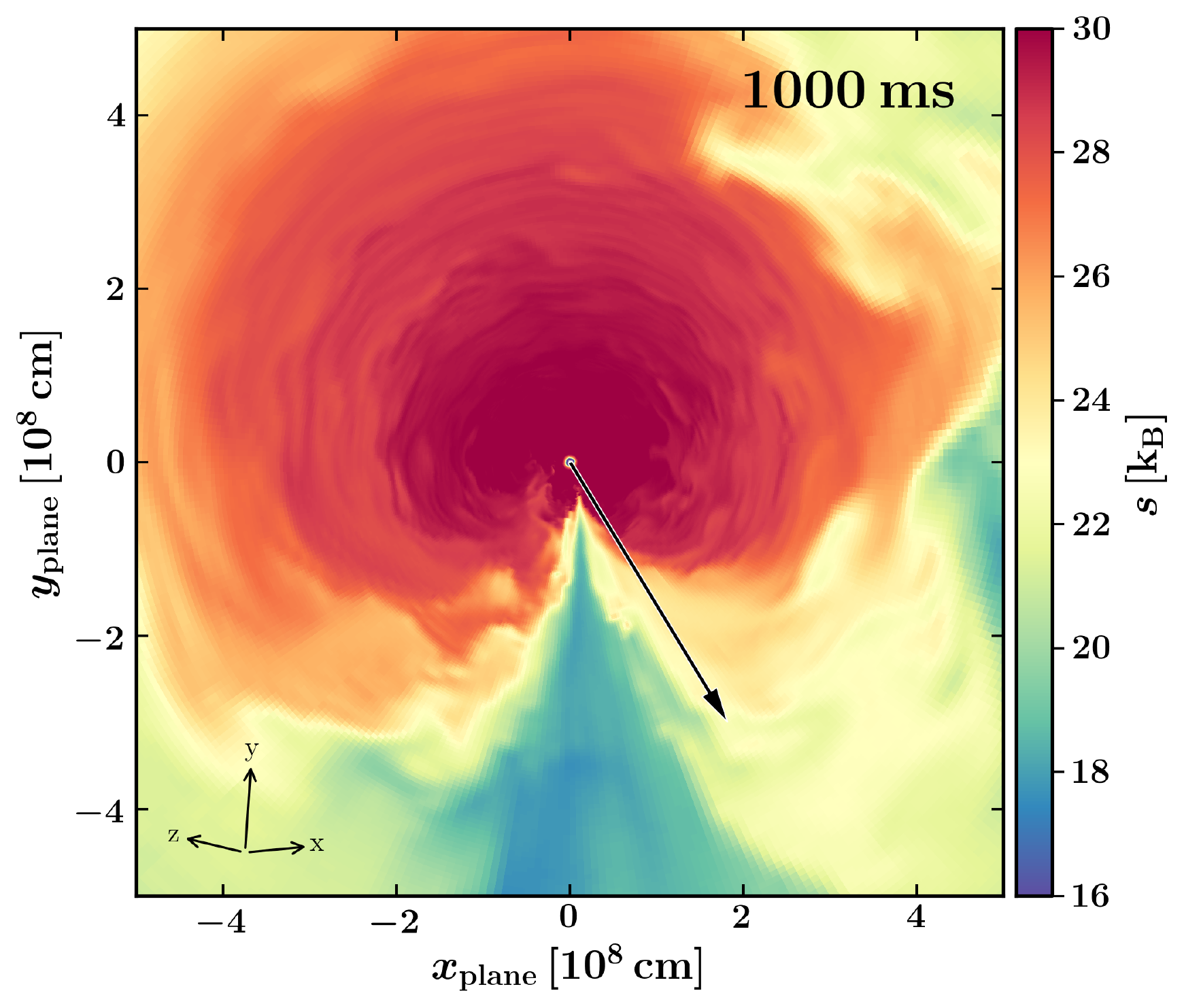}
\caption{Entropy per nucleon in slices displaying the late time evolution of
model O8-3D in continuation of Fig.~\ref{fig:3dsto}. The cross-sectional plane
is the same as in Fig.~\ref{fig:3dsto}.
\textit{Upper panels:} After 500\,ms, a strong downflow in the negative
$y$-direction penetrates inward and becomes even more prominent during the
following 250\,ms. \textit{Lower panels:} Final state of the simulation at
1\,s with a zoom into the central region shown in the \textit{bottom right panel}.
The final direction of the NS kick is in the hemisphere of the massive downflow
and nearly coincides with its position.}
\label{fig:3dsto_2}
\end{figure*}

Each of our five model sets consists of five simulation runs. In other
words, for each combination of values for our triple of parameters, we
performed five simulations using different seed perturbations.
One corresponding case per set was computed in 3D. Model 
set O5 is treated preferentially by repeating the 2D calculations for these
parameter values 20 times. The reason for this exception is that 
a recent detailed analysis of the Crab Nebula tends to favor a low blast-wave
energy (around $5\times 10^{49}$\,erg) for SN~1054 \citep{YangChevalier2015},
in line with fully relativistic 1D and 2D calculations of O-Ne-Mg core 
explosions \citep{vonGroote2014}.

All simulations were carried on for 1\,s of physical evolution, 
except for O16-3D, which could not be advanced beyond 750\,ms. At that
time this run terminated because of insufficient resolution at the shock at
very large distances from the center. Since this problem does not 
affect the NS-kick relevant inner ejecta and since all characteristic 
parameters of the explosion (in particular blast-wave energy, NS velocity
and acceleration) are basically converged, we accept the output at 
750\,ms as final result for model O16-3D.

Figure~\ref{fig:eexp} displays the time evolution of the diagnostic explosion
energies for all simulated 2D and 3D models. The variability 
for a fixed parameter set is caused by stochastic effects in the hydrodynamic
evolution as a result of the nonlinear growth of hydrodynamic instabilities
triggered by the initial random perturbations.
Moreover, Fig.~\ref{fig:eexp} introduces a color 
coding for the model of our five sets that will also be applied in 
Figs.~\ref{fig:hist} and \ref{fig:stat}. The diagnostic energy is defined
as the volume integral over the sum of internal, kinetic, and gravitational 
energy densities for all post-shock grid cells where this sum is 
positive.\footnote{In contrast to the explosion energy, the 
diagnostic energy
does not include the negative binding energy of the stellar layers ahead of
the outgoing SN shock. In the case of O-Ne-Mg cores, the surrounding
H/He envelope is so loosely bound (the absolute value of the binding energy 
is at most $\sim$10$^{47}$\,erg) and the shock expands so rapidly that the 
difference between diagnostic energy and explosion energy does not play any
role in practice when the diagnostic energy begins its rapid rise visible in
Fig.~\ref{fig:eexp}.}
Shortly after the SN shock has begun to accelerate outward, the latter 
condition is fulfilled for all expanding, high-entropy material in the layer
between shock front and gain radius, i.e., in the region where neutrino 
heating is responsible for net energy deposition. Once the shock has
reached a radius of several 1000\,km, the diagnostic energy is essentially 
equal to the (instantaneous) explosion energy, because the gravitational 
binding energy of the overlying stellar layers ahead of the shock becomes
extremely low in the considered progenitor 
\citep[see figure~1 in][]{Radiceetal2017}.
Since the final explosion energy is basically reached after $\sim$300\,ms 
at the latest, we will not distinguish between diagnostic energy and 
explosion energy in the remainder of our paper.

\subsection{Explosion properties of ECSNe}
\label{sec:ex_prop}
Our simulations exhibit an explosion behavior very similar to the published
ECSN models discussed in the literature 
(see \citealt{Kitauraetal2006,Huedepohletal2010,Fischeretal2010}
for 1D simulations and \citealt{Jankaetal2008,Wanajoetal2011,vonGroote2014,Radiceetal2017} 
for 2D results).
We therefore summarize here only the basic features and focus specifically
on those aspects that are of interest in the context of NS kicks to be discussed
in the subsequent sections.

The extremely steep density gradient at the edge of the O-Ne-Mg core crucially
determines the shock dynamics in ECNSe. 
In response to the rapidly decreasing ram pressure of the infalling
matter when the shock reaches the core/envelope boundary, fast outward shock 
acceleration sets in after less than $\sim$100\,ms of slower, but continuous,
shock expansion 
\citep[very similar to the evolution shown in figure~3 of][]{Jankaetal2008}.
Within this time, nonradial asymmetries begin to develop only in the close
vicinity of the NS, because neutrino-heated high-entropy matter outside 
of the gain radius becomes buoyant and starts rising in Rayleigh-Taylor plumes.
This phenomenon takes place basically independently of the explosion energy and
of the dimension of the models (see Fig.~\ref{fig:2dsto}, upper row, and
Fig.~\ref{fig:3dsto}, upper left panel). When the most extended of these 
convective bubbles reach radii of a few hundred kilometers at $\sim$100\,ms, the
shock has already propagated to more than $10^4$\,km 
\citep[see figure~3 in][]{Jankaetal2008}
and is not affected by the hydrodynamic instabilities growing far behind in its wake.
An almost spherical expansion of the shock is therefore a generic feature of ECSNe.

Figure~\ref{fig:2dsto} shows snapshots of the 2D entropy distribution at different 
times between 100\,ms and 1\,s for the axisymmetric models O5-vl (left column) 
and O16-vl (right column), which
represent a low-energy and a high-energy explosion, respectively.
Both models exhibit an almost self-similar expansion of the Rayleigh-Taylor plumes
that have developed already until 100\,ms. The expansion is slightly faster for
the more than three times more energetic explosion of model O16-vl.

The lower-energy model O5-vl and the higher-energy model O16-vl are representatives
of two antipodes with respect to their behaviors. In O16-vl the neutrino luminosities
from the PNS are high enough to drive a powerful baryonic outflow 
(``neutrino wind'') by neutrino-energy deposition near the PNS surface.
This essentially spherically symmetric mass outflow leads to the continuous,
slow growth of the explosion energy visible in Fig.~\ref{fig:eexp} from
$\sim$400\,ms until the end of the simulation. During this phase
the high-entropy wind region can be seen as red sphere
around the center in the right panels of Fig.~\ref{fig:2dsto}. Because the wind
fills the central volume, the Rayleigh-Taylor plumes of the convective layer
are pushed outward and exhibit signs of a growing compression in the radial 
direction. The two distinct downflows at the poles (with the stronger one at the
north pole along the $+z$-direction) are axis artifacts fostered by the flow
constraints connected to the assumption of axisymmetry.

\begin{figure*}[!]
\centering
 \includegraphics[width=0.32\textwidth]{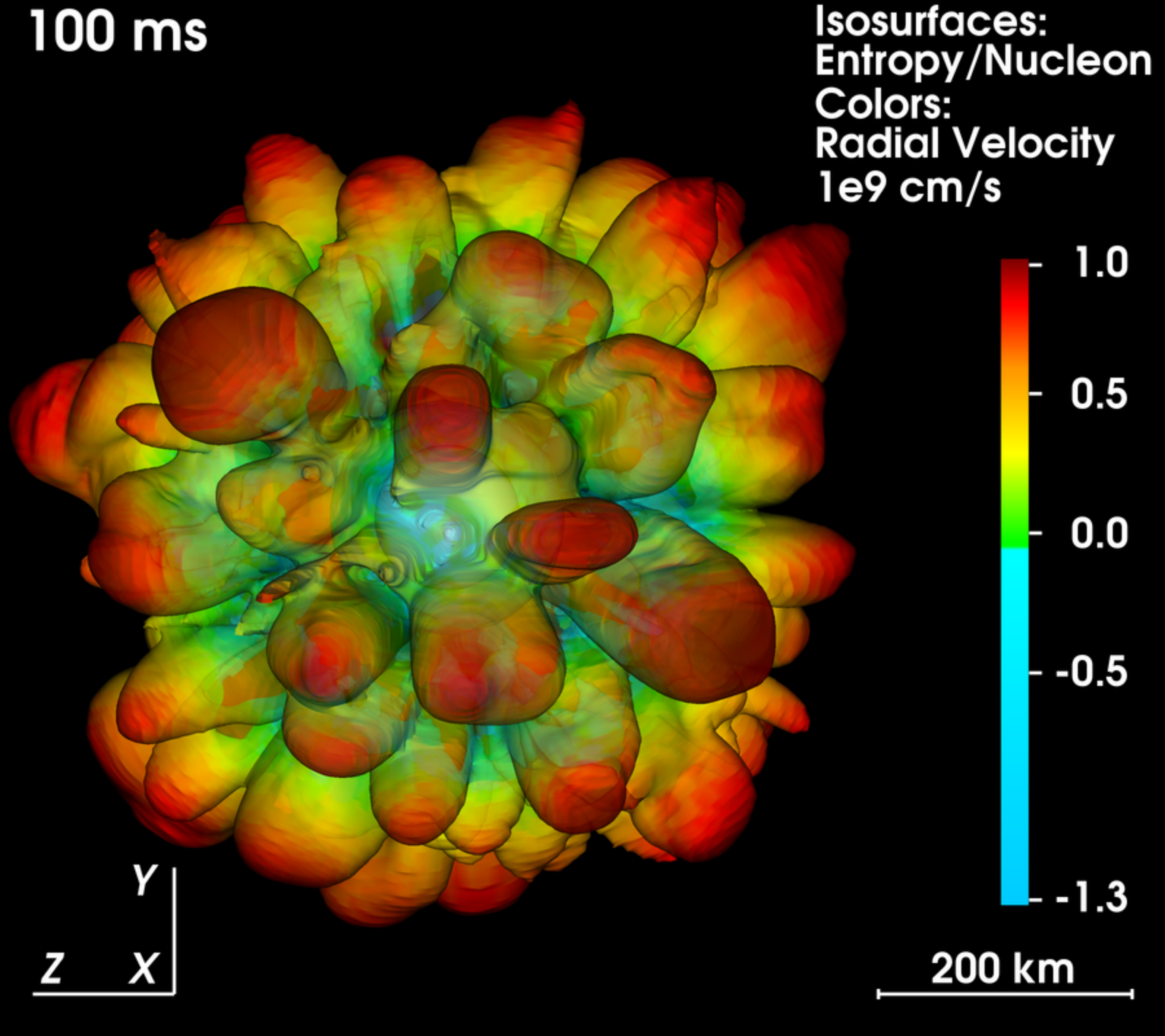}
 \hfill
 \includegraphics[width=0.32\textwidth]{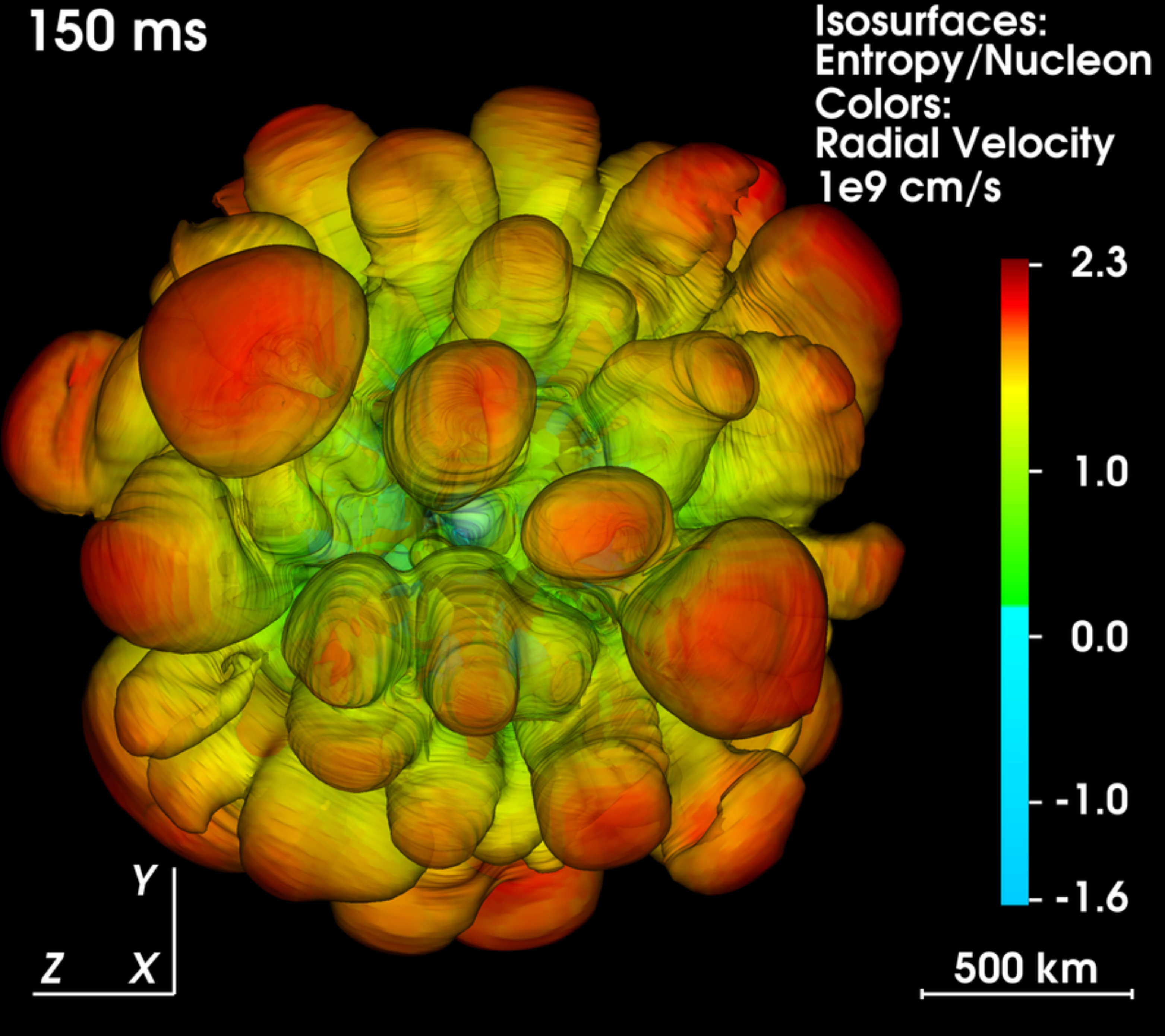}
 \hfill
 \includegraphics[width=0.32\textwidth]{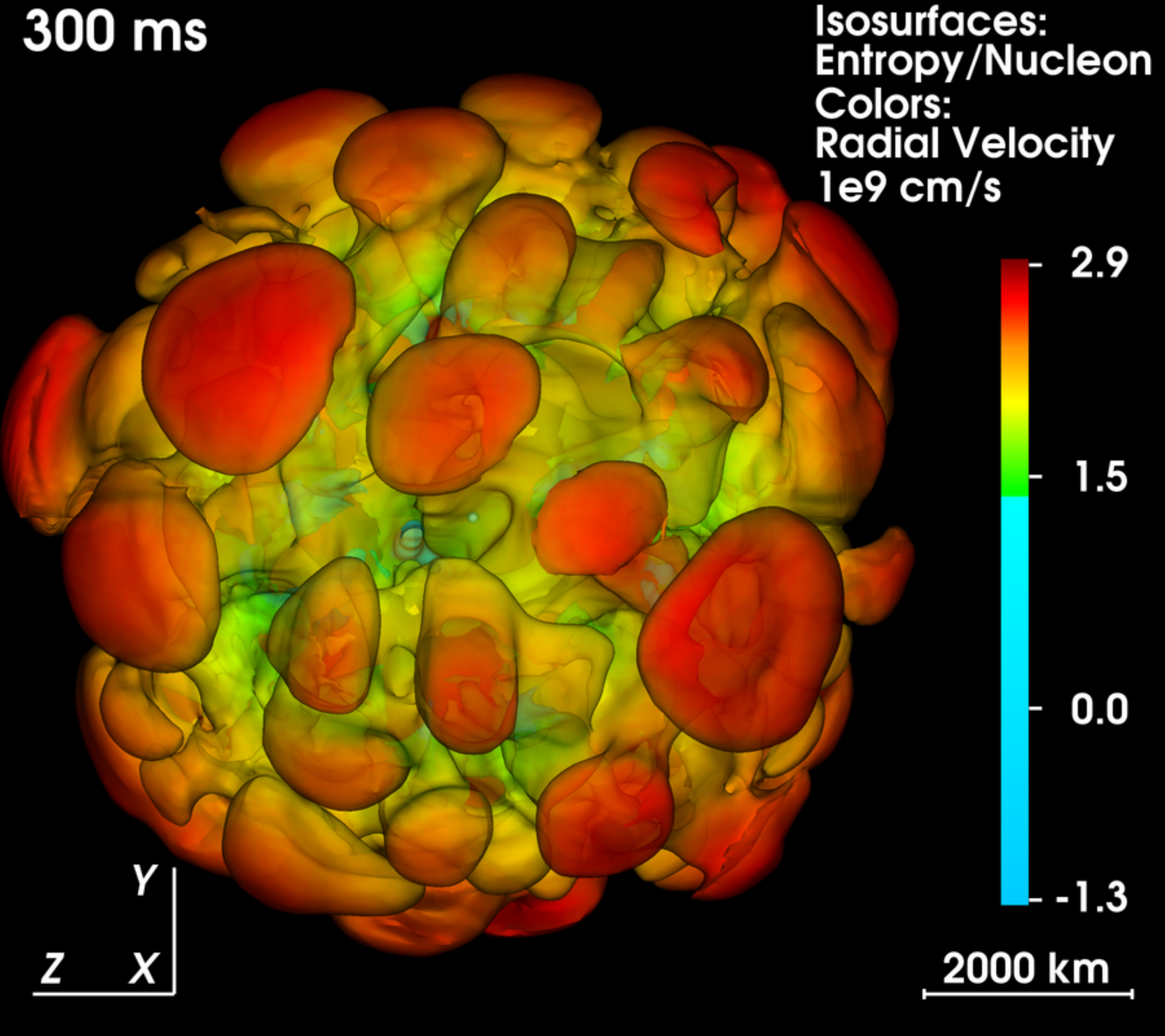}\\
 \vspace{3mm}
 \includegraphics[width=0.32\textwidth]{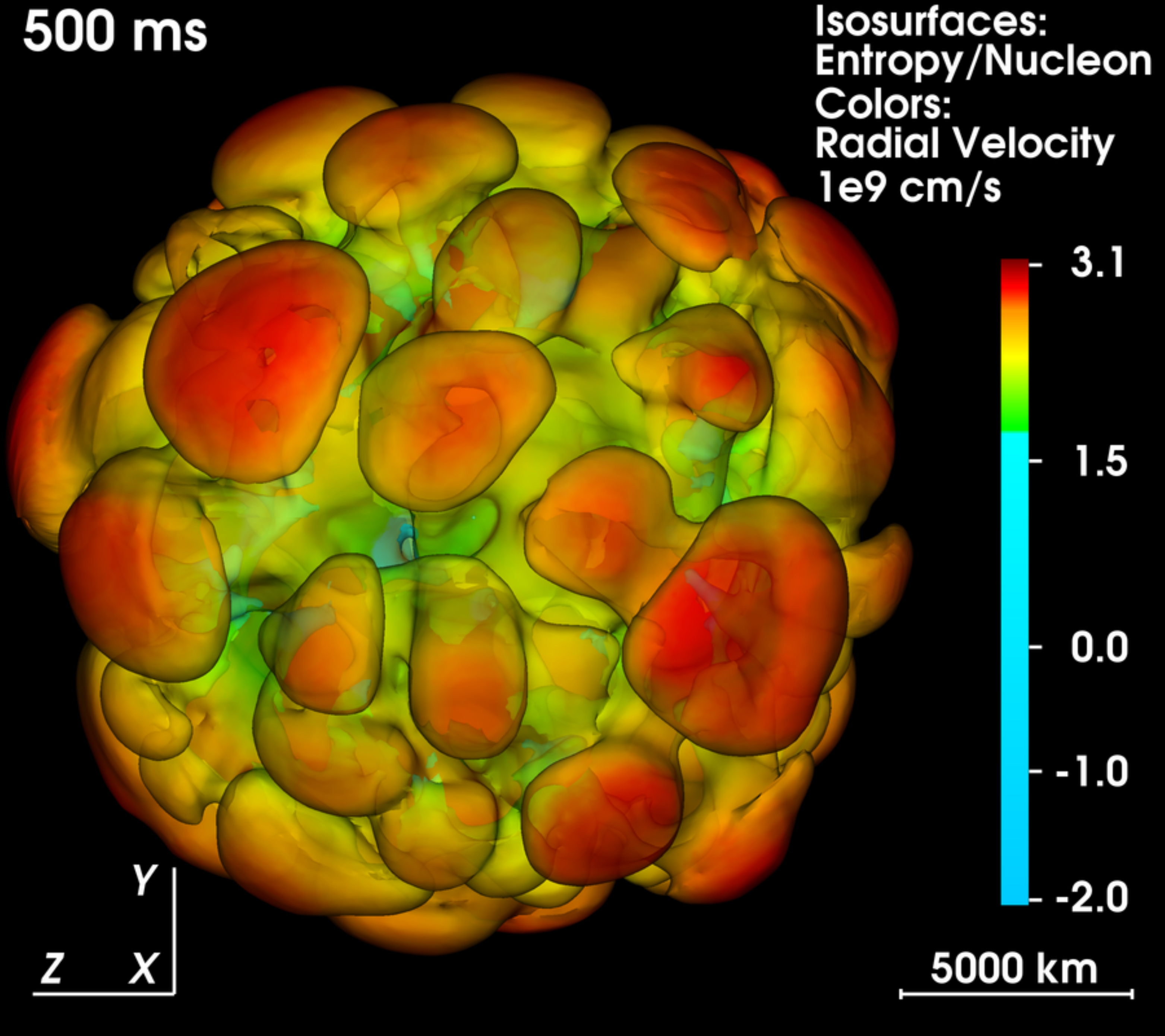}
 \hfill
 \includegraphics[width=0.32\textwidth]{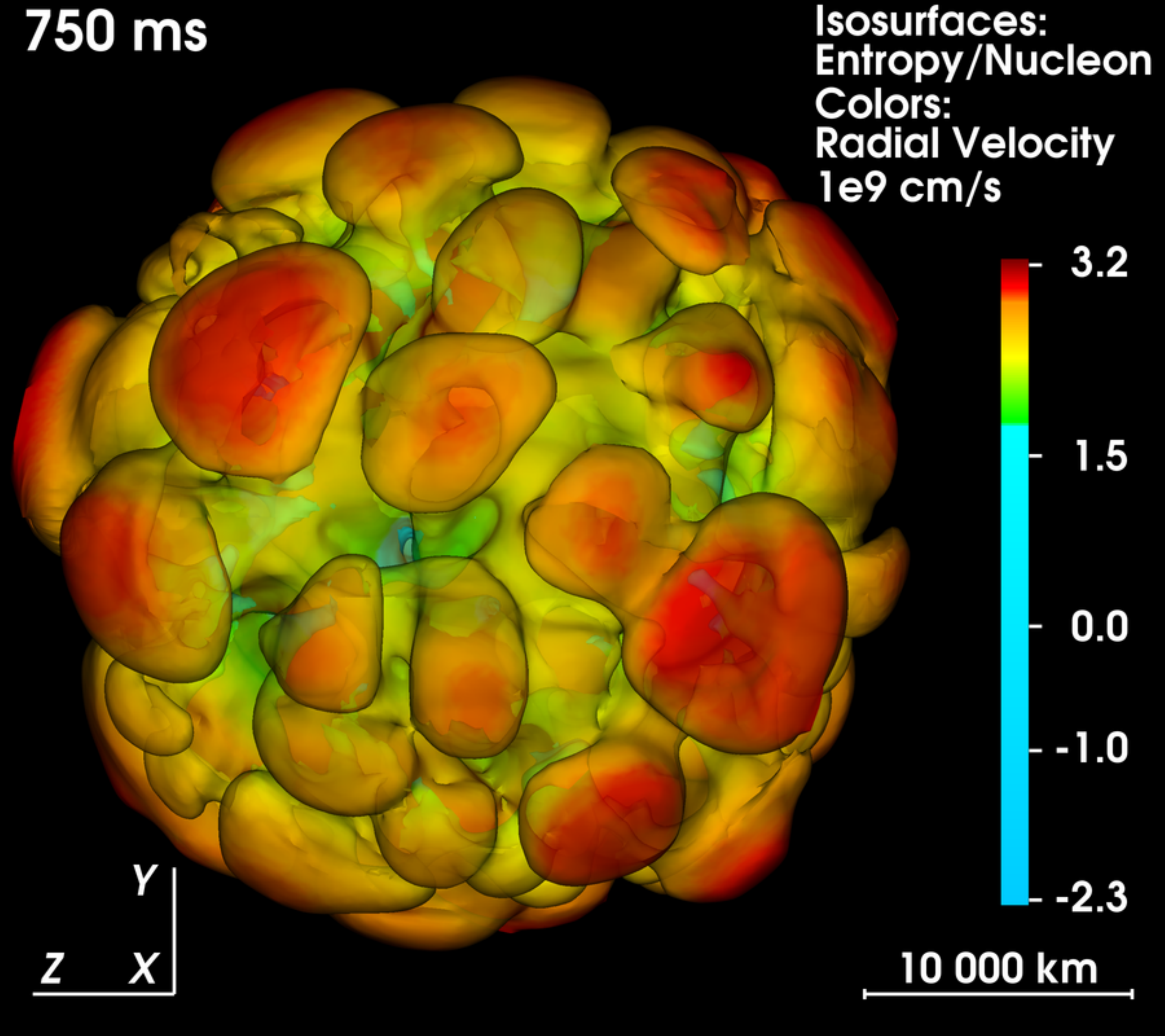}
 \hfill
 \includegraphics[width=0.32\textwidth]{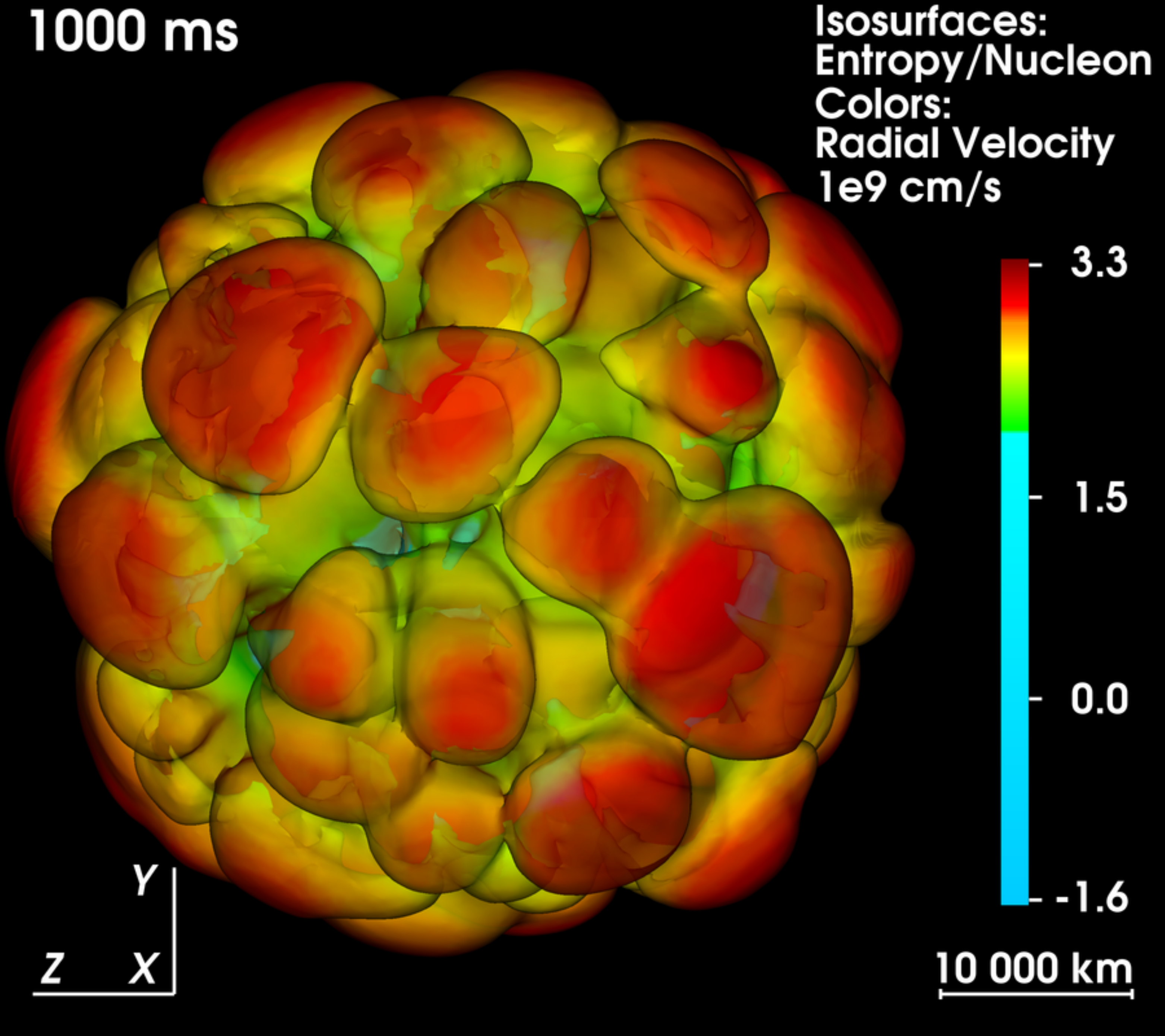}
\caption{Three-dimensional visualization of model O8-3D at 100, 150, 300, 500, 750,
and 1000\,ms (\textit{from top left to bottom right}). The images show isoentropy
surfaces (for suitably chosen values of the entropy per nucleon), color coded with
the radial velocity in units of $10^9$\,cm\,s$^{-1}$ as given by the color bar.
Rayleigh-Taylor plumes of buoyant, neutrino-heated matter expand almost 
self-similarly (the growing spatial scale is indicated by the yardstick in the
lower right corner of each panel), and low-order multipolar structures are
insignificant.}
\label{fig:elena_plots}
\end{figure*}

In contrast, the neutrino heating by the lower luminosities in model O5-vl is too
weak to shed off matter from the PNS surface in a strong neutrino-driven wind. 
Therefore the explosion energy levels off to its final value after $\sim$300\,ms
(Fig.~\ref{fig:eexp}), and convective filaments keep on billowing around the NS
until the end of the simulation. The initial Rayleigh-Taylor plumes continue to
expand in a self-similar manner.

Figures~\ref{fig:3dsto} and \ref{fig:3dsto_2} depict the entropy per nucleon, 
color coded, at different times in a cross-sectional plane of the 3D data
from the evolution of model O8-3D. This model develops the largest ejecta 
asymmetry and the highest NS kick of all of our 3D runs. 
The angular scale of the Rayleigh-Taylor plumes
is corroboratively similar to the 2D cases displayed in Fig.~\ref{fig:2dsto}
and clearly dominated by higher-order spherical harmonics and nearly
self-similar expansion already after 150\,ms (see Fig.~\ref{fig:elena_plots}).
The sequence of plots in Fig.~\ref{fig:3dsto_2}
visualizes the formation of a prominent, long-lasting
downflow, which grows after $\sim$400\,ms from a massive pocket of cool, 
low-entropy material enclosed by the outward-rising bubbles. 
While the 
development of buoyant Rayleigh-Taylor plumes of neutrino-heated matter with
similar geometrical patterns is a generic feature of all of our 3D models, 
the long-lasting downflow is specific to model O8-3D.
This downflow determines the final direction and magnitude of the NS kick,
which is generated by the gravitational tug-boat mechanism 
and largest in model O8-3D compared to the other 3D cases, 
to be discussed in the next section.

\section{Neutron star kicks in ECSNe}
\label{sec:kicks}
\subsection{Neutron star kicks -- physics}
\label{sec:kickphysics}
The gravitational tug-boat mechanism attributes natal kicks of NSs to
anisotropic ejection of matter that exerts gravitational and hydrodynamic forces 
on the NS \citep{Schecketal2006,Wongwathanaratetal2010,Wongwathanaratetal2013,Janka2013,Janka2017}.
In our simulations, the NS is tied to the center of the numerical grid 
because of the excluded core volume, which is replaced by a gravitating point mass 
and an inner grid boundary.
However, due to the asymmetries growing by nonradial hydrodynamic instabilities
at the beginning of the SN explosion, the ejecta can acquire a net linear
momentum on the grid. The corresponding recoil velocity of the NS,
$\vect{v}_\mathrm{ns}$, can be evaluated by applying linear momentum conservation 
of the ejected gas plus the compact remnant, i.e., the momentum carried by the NS
is given by the negative of the ejecta momentum $\vect{P}_\mathrm{gas}$. 
This yields
\begin{equation}
 \vect{v}_{\mathrm{ns}}(t) = -\frac{\vect{P}_\mathrm{gas}(t)}{M_\mathrm{ns}(t)}\,,
\label{eq:momcon}
\end{equation}
where the mass of the NS is given by $M_{\mathrm{ns}}$, which we take to be
the baryonic mass enclosed 
by the NS radius $R_{\mathrm{ns}}$. The latter is defined as the (angle dependent)
radius where the density drops below $10^{11}\,\mathrm{g\,cm}^{-3}$.
The gas momentum $\vect{P}_\mathrm{gas} = \int_{R_\mathrm{ns}}^{R_\mathrm{ob}}\rho
\vect{v}\,\diff V$ is the net linear momentum on the computational 
grid exterior to the NS.
It should be kept in mind that this momentum is exclusively associated with the 
ejecta mass, $M_\mathrm{ej} = \int_{R_\mathrm{ns}}^{R_\mathrm{s}}\rho\,\diff V$
(with $R_\mathrm{s}$ being the angle-dependent shock radius), because the linear
momentum of the gas ahead of the shock is negligible (since the shock is nearly
spherical around the grid center and the star is spherical and basically
at rest). 

In order to better understand the exact acceleration mechanism acting on the NS
and to back up the direct numerical evaluation of the simulation 
results by a more detailed physical consideration
\citep[following previous works for NS kicks in SNe from Fe-core progenitors;][]{Schecketal2006,Wongwathanaratetal2010,Wongwathanaratetal2013,Nordhausetal2010,Muelleretal2017}, 
we separate the individual forces that act on the NS and cause its kick
\citep[see][]{Schecketal2006}:
\begin{eqnarray}
\!\!\!\!\!\!\! \frac{\diff}{\diff t}\vect{P}_\mathrm{ns} &=&  
\frac{\diff}{\diff t}\left(M_\mathrm{ns}\,\vect{v}_\mathrm{ns}\right) \nonumber \\
&=& -\!\!\! \oint\displaylimits_{r=r_0}\!\! p\,\diff\vect{S} 
- \!\!\! \oint\displaylimits_{r=r_0}\!\! \rho\vect{v}v_r \,\diff S + 
\!\!\! \int\displaylimits_{r>r_0}\!\! GM_\mathrm{ns}\frac{\vect{r}}{r^3}\,\diff m \label{eq:forces} \\
&\approx& M_\mathrm{ns}\, \frac{\diff \vect{v}_\mathrm{ns}}{\diff t} \,,
\label{eq:forces2}
\end{eqnarray}
where $\diff S = r^2 \sin\theta\diff\theta\diff\phi$ is the surface element on
a sphere of radius $r$. Using the approximative relation in the
third line of this equation 
(and thus adopting the analysis of NS kicks in Fe-core SNe), 
dividing by $M_\mathrm{ns}$, and integrating over time
yields the NS recoil velocity, provided $M_\mathrm{ns}$ does not change with 
time. For the calculation of NS kicks in SNe of Fe-core progenitors this 
turned out to work very well, because in those cases the crucial phase of
NS acceleration occurs after the onset of the SN explosion and continues
for several seconds, see 
\citet{Schecketal2006,Wongwathanaratetal2010,Wongwathanaratetal2013,Nordhausetal2010,Muelleretal2017}. 
During this phase of evolution, the assumption of a constant PNS
mass is good roughly on the percent level.
In the case of O-Ne-Mg core explosions considered here, however, the NS kicks 
happen so fast and early that variations of the NS mass might have
a more relevant impact. We will discuss this in Sect.~\ref{sec:kickresults}.

The first term on the r.h.s.\ of Eq.~(\ref{eq:forces}) accounts for pressure 
variations over a spherical surface chosen at $r_0 = 1.3 R_\mathrm{ns}$ 
\citep{Wongwathanaratetal2013}. The momentum flux across this surface (second 
term) can be decomposed into down- and outflows, depending on the sign of the 
radial velocity $v_r$. Both terms only contribute significantly if anisotropically
distributed matter is in sonic contact with the sphere of evaluation.
Otherwise, if asymmetries are present only in matter that is hydrodynamically 
detached from the NS, the only remaining possibility to accelerate the NS is
by the long-range forces of gravity, which are accounted for by the third term.
In contrast to the hydrodynamic forces, the NS interacts gravitationally with
matter in the entire volume under consideration. Therefore the third term
typically constitutes the 
most long-lasting contribution to the total acceleration. As the gravitational 
force drops with the square of the distance, material that stays longer in the
vicinity of the NS has a larger accelerating influence on it. This so-called 
gravitational tug-boat mechanism \citep{Wongwathanaratetal2013} pulls the NS
into the direction of the most massive, most slowly moving ejecta. These ejecta 
preferentially
lie in the direction where the explosion is weaker. This means that the NS is 
kicked opposite to the hemisphere of the stronger explosion as expected for a
hydrodynamical acceleration associated with blast-wave asymmetries.

\begin{figure*}[!]
\centering
\includegraphics[width=0.49\hsize]{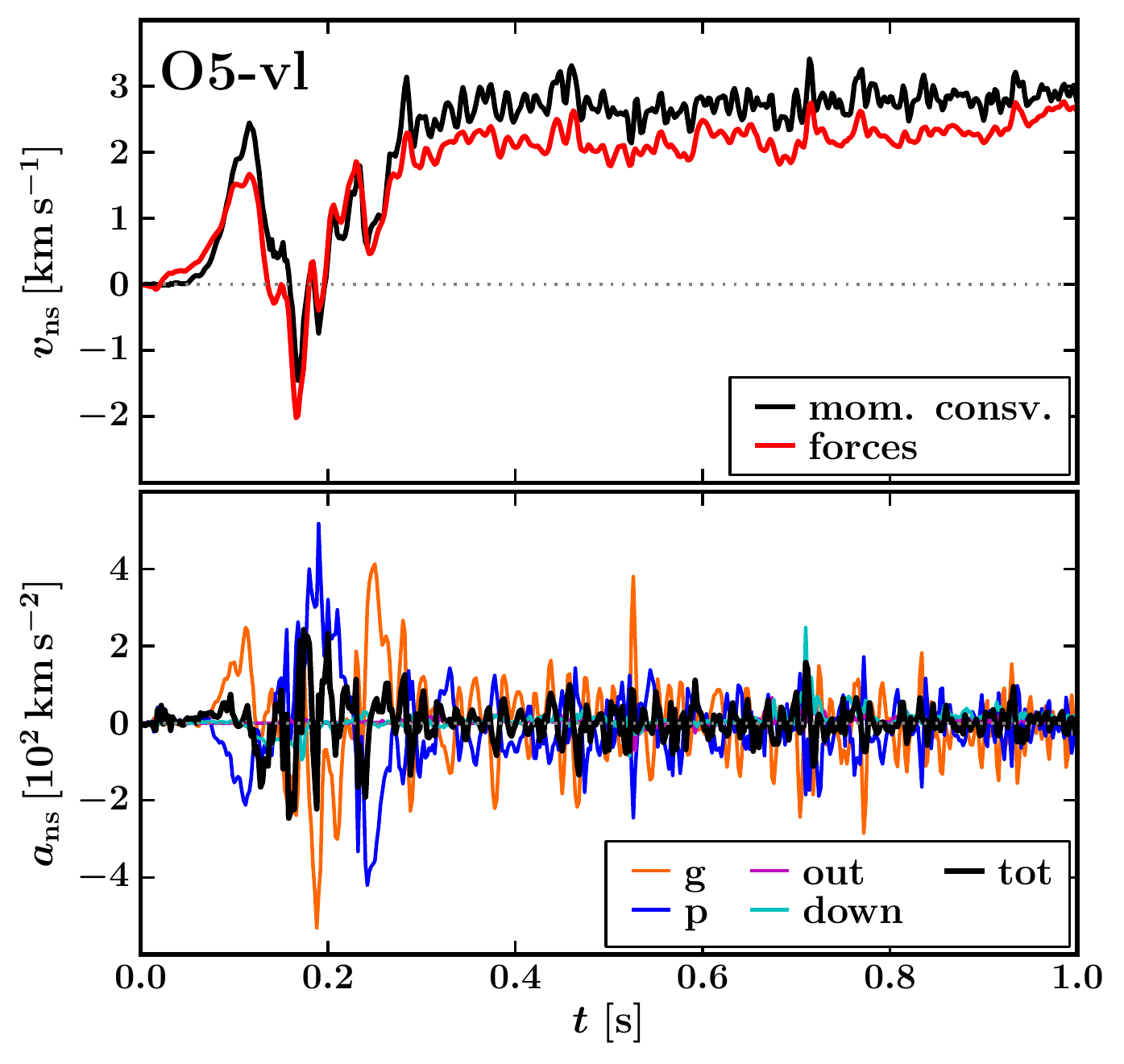}
\hfill
\includegraphics[width=0.49\hsize]{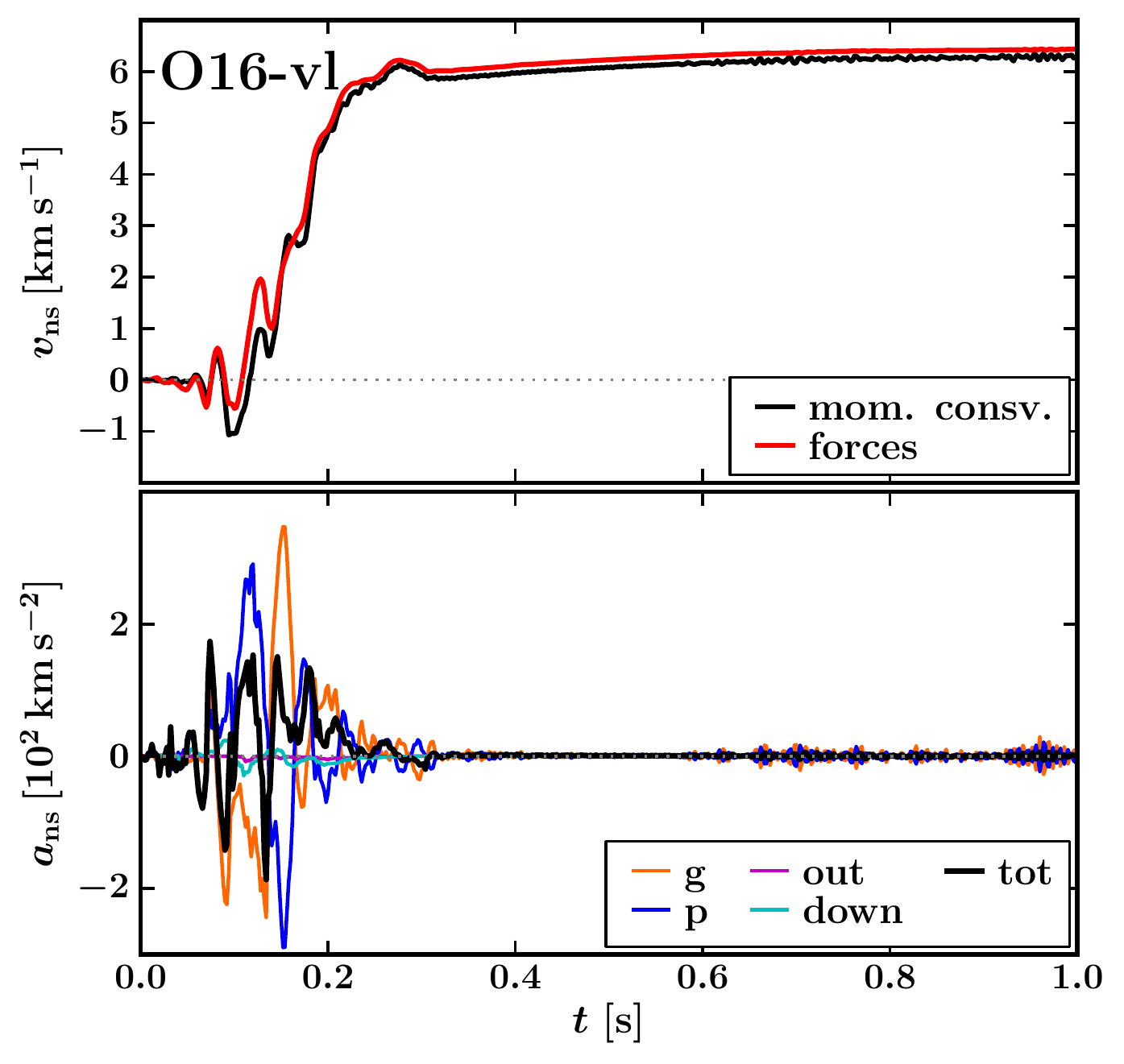}
\caption{Time evolution of the NS kick velocity and of the different
contributions to the NS acceleration for the 2D models of
Fig.~\ref{fig:2dsto}, O5-vl (\emph{left}) and O16-vl (\emph{right}).
\emph{Upper panels:} Evolution of the kick velocity computed
from the assumption of momentum conservation during the hydrodynamical
simulation (i.e., the net momentum of ejected gas is assumed to be equal
and opposite to the NS momentum; \textit{black line}) and by direct time
integration of the acceleration through the sum of all contributing forces
(\textit{red line}; see Sect.~\ref{sec:examplarycases} 
for details). Despite considerably different explosion energies and
ejecta dynamics of the two models (Fig.~\ref{fig:2dsto}), the NS in both
cases has reached its terminal velocity at the end of the simulations.
\emph{Lower panels:} Individual components of the acceleration (specific
forces) acting on the NS with ``g'' for gravitation, ``p'' for pressure,
and ``out'' and ``down'' for the acceleration associated with momentum
transfer due to outflows and downflows, respectively; ``tot'' denotes the
sum of all individual contributions to the NS acceleration.}
\label{fig:2dkick}
\end{figure*}

\begin{figure*}[!]
\centering
\includegraphics[width=0.8\textwidth]{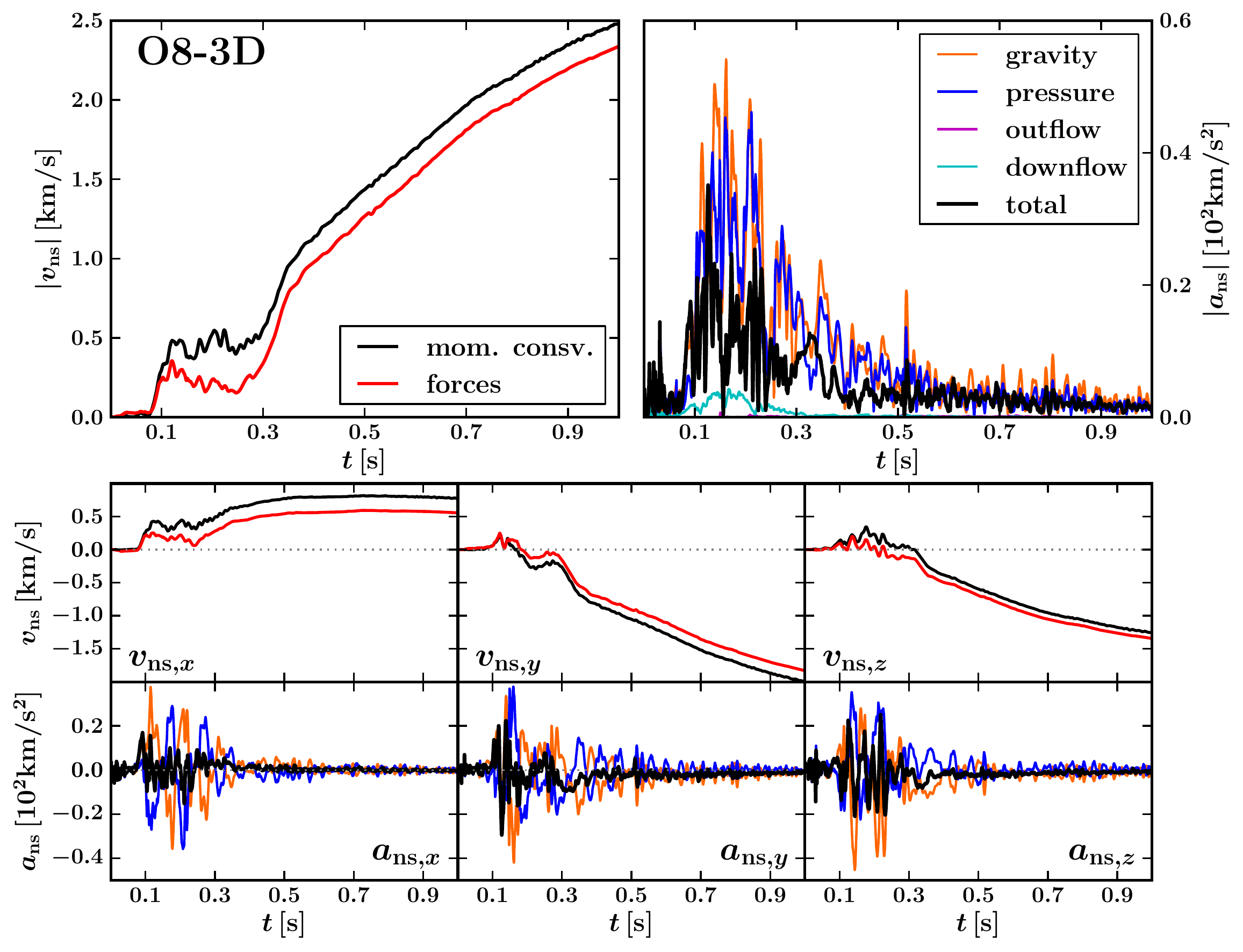}
\caption{\textit{Upper panels:} Time evolution of the absolute values of
the NS kick velocity (\textit{left}) and of the different contributions
to the NS acceleration
(\textit{right}) for the 3D model O8-3D of Figs.~\ref{fig:3dsto} and
\ref{fig:3dsto_2} (color coding and image elements analogous to
Fig.~\ref{fig:2dkick}). It is noteworthy that the kick velocity has not yet
reached its terminal value by the time the simulation is stopped. The
reason is the long-lasting, massive downflow near the $(-y)$-direction
visible in Fig.~\ref{fig:3dsto_2}. This is confirmed by the \textit{lower panels},
which show the evolution of the NS kick velocity (\textit{top row}) and
acceleration (\textit{bottom row}) in the different coordinate directions.
At the end of the simulation the $y$-component of the velocity vector
exhibits the most significant growth to negative values. The acceleration
components perform fluctuations with amplitudes and variation frequencies
very similar to those in the 2D models.}
\label{fig:3dkick}
\end{figure*}

\subsection{Results}
\label{sec:kickresults}
\subsubsection{Exemplary cases}
\label{sec:examplarycases}
With insight gained about the forces that contribute to the acceleration 
of the NS, we return to the exemplary 2D models O5-vl and O16-vl shown 
in Fig.~\ref{fig:2dsto}. In such calculations the assumed axisymmetry
constrains the NS kick direction to coincide with the symmetry axis.
For both models we obtain a positive kick velocity, i.e., a kick vector
pointing to the $(+z)$-direction (see Table~\ref{tab:kick} for results of
all models).
This complies with the observation that both models develop a stronger 
downflow at the north pole, indicating the hemisphere where the SN blast
is weaker. While O16-vl is on the extreme side with its NS kick velocity
of more than 6\,km\,s$^{-1}$, model O5-vl with $\sim$3\,km\,s$^{-1}$ at
the end of our simulation is a fairly average 2D case.

Because of the development of one distinct downflow during the later
stages ($t\gtrsim$400\,ms) of the computed evolution, our exemplary 3D model
O8-3D (Figs.~\ref{fig:3dsto} and \ref{fig:3dsto_2}) exhibits a kick geometry
very similar to the discussed 2D cases. The kick direction almost coincides
with the downflow, indicating that
this downflow is the main feature that determines the NS kick
direction and magnitude. The deviation from a perfect alignment is 
caused by the massive ejecta clumps that expand with relatively lower
velocities in the right hemisphere (yellow regions with intermediate entropy 
values and relatively high densities at the one o'clock and five o'clock
positions in the bottom right panel of 
Fig.~\ref{fig:3dsto_2}). Model O8-3D with its prominent single downflow,
however, is not representative for our sample of computed 3D cases. Its
NS kick velocity of 2.5\,km\,s$^{-1}$, which is still growing with
a comparatively high rate of 1.2 \,km\,s$^{-2}$ at one second, is the
largest of all 3D models. This suggests that the ejecta geometry that
has formed in model O8-3D is not a generic feature in 3D,
as mentioned before (Sect.~\ref{sec:ex_prop}). 
The relation of explosion asymmetries and NS kicks in 2D and 3D
models with be further discussed in Sects.~\ref{sec:compilation} 
and~\ref{sec:auk}.

Figures~\ref{fig:2dkick} and \ref{fig:3dkick} display the time evolution 
of the NS velocities and the different contributions to the NS 
acceleration computed with Eqs.~(\ref{eq:momcon}) and 
(\ref{eq:forces},\ref{eq:forces2}) 
for the models under consideration. In Fig.~\ref{fig:2dkick}
positive values correspond to motion/acceleration in 
the $(+z)$-direction of the 2D polar grid. The NS velocities computed 
directly from the hydrodynamical results with Eq.~(\ref{eq:momcon}) 
under the assumption of momentum conservation (upper panels; black lines)
agree very well with the summed effects of all accelerating forces as obtained
by an analysis using Eq.~(\ref{eq:forces}) (upper panels; red lines). 
The less energetic explosion of model O5-vl features stronger accelerations,
and the NS velocity exhibits more dramatic changes between positive and 
negative values during the initial $\sim$200\,ms than in the higher-energy
explosion of model O16-vl, where the NS velocity is fairly monotonically 
boosted to its final value. In both cases the terminal speed of the NS is
reached within roughly 300\,ms. But in model O5-vl the NS motion continues
to exhibit low-amplitude fluctuations caused by the persisting nonradial
flows in the low-density medium around the NS (see Fig.~\ref{fig:2dsto},
left column), which lead to variations of the acceleration whose amplitudes
decrease only slowly with time. 

In model O8-3D (Fig.~\ref{fig:3dkick}), the
magnitude and direction of $\vect{v}_\mathrm{ns}$ have not yet asymptoted
by the end of the simulation. However, the NS acceleration decreases
continuously so that a tendency towards saturation is clearly visible and
a further growth of the NS kick by at most $\sim$0.5--1\,km\,s$^{-1}$ may
be expected. The velocity components in the $(-y)$ and $(-z)$-directions 
dominate, as may be concluded from an inspection of the orientation of the
kick vector relative to the tripod in Fig.~\ref{fig:3dsto_2}.

The dominant contributions to the NS acceleration result from gravitational
and pressure forces (orange and blue lines, respectively, in the acceleration 
panels of Figs.~\ref{fig:2dkick} and \ref{fig:3dkick}), which essentially
always counteract and nearly balance. Since pressure forces require
matter to be in sonic contact with the NS surface, the corresponding 
acceleration is exerted by mass in the very close vicinity of the NS. Gravity
causing the almost same effect as pressure forces, but with opposite sign,
suggests that also the acceleration by gravity forces is mostly associated
with matter in the near surroundings of the NS in our models. That is, the
same clumps of matter that pull the NS gravitationally also push it
away by means of pressure. Because of the extremely rapid outward expansion
of the ejecta in ECSNe and the lack of a large-scale dipolar asymmetry, the 
gravitational influence of outflowing gas quickly abates when this
material rushes to larger
radial distances. This explains why also the gravitational influence on the
NS is dominantly linked to nearby matter.

Hydrodynamic mass flows towards as well as away from the NS always 
play only an ancillary role for the acceleration of the NS. The
medium surrounding the NS has a fairly low density, and powerful accretion
flows with supersonic velocities (like in explosions of more massive 
progenitors) are absent. For this reason the momentum fluxes associated
with anisotropic flows around the NS are very small. The very fast 
expansion of the asymmetric ejecta and the rapid decline of the densities 
around the PNS in ECSNe also explain why the NS acceleration is finished
within only $\sim$300\,ms, i.e., within a much shorter timescale than
in massive Fe-core progenitors, where the NS experiences an acceleration
over several seconds \citep{Schecketal2006,Wongwathanaratetal2013,Muelleretal2017}.

Note that minor differences between the NS velocities determined with
Eqs.~(\ref{eq:momcon}) and (\ref{eq:forces}) in the cases of models O5-vl
and O8-3D (black and and red lines in the panels of Figs.~\ref{fig:2dkick}
and \ref{fig:3dkick} showing $v_\mathrm{ns}$ versus $t$) 
develop already at an early stage of the evolution (100--300\,ms p.b.) and 
stay basically constant later on. These differences can be 
understood by two facts. First, during the early post-bounce phase
mass inflows and outflows as well as gravitational settling
can still lead to small changes of the PNS mass.\footnote{The reader 
is reminded that we define the PNS mass, 
$M_\mathrm{ns}(t)$, somewhat arbitrarily as the mass above a density of 
$10^{11}$\,g\,cm$^{-3}$ at time $t$. This mass can increase as the PNS
contracts and settles and accretes, or it can decrease when neutrino 
heating blows matter off the PNS surface.} 
However, when computing $\vect{v}_\mathrm{ns}(t)$ by integrating 
Eq.~(\ref{eq:forces}), we take into account only the first term in the 
expression $\diff\vect{P}_\mathrm{ns}/\diff t =
M_\mathrm{ns}\diff\vect{v}_\mathrm{ns}/\diff t +
\vect{v}_\mathrm{ns}\diff M_\mathrm{ns}/\diff t$ but ignore the second
term (which leads to the approximation of Eq.~\ref{eq:forces2}). This
implies an error of order $\Delta M_\mathrm{ns}/M_\mathrm{ns}$ in the 
corresponding estimate of the NS velocity, when $\Delta M_\mathrm{ns}$
is the change of the NS mass during the phase of NS acceleration.
In our O-Ne-Mg core explosions, $M_\mathrm{ns}$ varies by less than
$\sim$5\% at $t\gtrsim 100$\,ms post bounce and by less than 
$\sim$1\% at $t\gtrsim 200$\,ms
post bounce. For this reason the omission of the second term of
$\diff\vect{P}_\mathrm{ns}/\diff t$ in Eq.~(\ref{eq:forces2}) in
combination with Eq.~(\ref{eq:forces}) can cause errors only of
order $\sim$1--5\% in $|\vect{v}_\mathrm{ns}|$, 
and this only at $t \lesssim 200$\,ms after bounce.
The main reason for discrepancies larger than that between results 
from Eqs.~(\ref{eq:momcon}) and (\ref{eq:forces}), in particular at 
$t \gtrsim 200$\,ms after bounce, are numerical integration errors
associated with the discrete time sampling 
of the simulation outputs that are subsequently
post-processed for evaluating the different terms in
Eq.~(\ref{eq:forces}). When the NS kick develops in a phase of very
rapid variations, the finite time sampling in steps of typically
about 0.5--1\,ms can be too coarse to capture all spikes in the 
acceleration terms. 
Usually the associated errors are of minor relevance when
the net NS kicks are large or temporal variations of the acceleration
terms happen on relatively long timescales. In the present models,
however, the different contributions to the NS acceleration 
fluctuate rapidly and mostly cancel 
each other, in particular when the net kicks become very small.
This is the case, e.g., in the left panels of Fig.~\ref{fig:2dkick} and
in Fig.~\ref{fig:3dkick}, which display cases with relatively big 
deviations between results from momentum conservation in the 
hydrodynamic simulation (Eq.~\ref{eq:momcon}; black lines) and results
from post-processing of the accelerating forces (Eq.~\ref{eq:forces};
red lines). Despite the asymptotic quantitative differences on the level
of 10--20\% in these two cases, however, both the black and red lines 
still follow each other closely and in the right panels of 
Fig.~\ref{fig:2dkick} even match quantitatively. We therefore deem
the sum of the force terms considered in Eq.~(\ref{eq:forces}) to be
a satisfactory confirmation for the NS kicks deduced
from our hydrodynamic simulations.

A second note at this place concerns a possible neutrino-induced kick.
Evaluating our models for anisotropic neutrino emission, we obtain 
only tiny effects, because the lack of high-density layers
around the degenerate progenitor core prevents the occurrence of massive
accretion flows to the NS, at least according to our explosion models (and 
understanding of the explosion mechanism) and for current ECSN progenitors.
Even in massive iron-core progenitors,
neutrino-induced kicks due to anisotropic neutrino emission from asymmetric
accretion flows are usually very small compared to the hydrodynamical kicks 
mediated by the gravitational tug-boat mechanism. Corresponding results
and the reasons for this finding are presented and explained in
\citet{Schecketal2006} and \citet{Wongwathanaratetal2013}. 
However, a possible NS kick
by anisotropic transport of neutrinos out of the NS interior cannot
be evaluated in our simulations, because the NS core
is excised from the computational domain and the NS evolution is not 
treated in detail. Instead, spherically symmetric neutrino luminosities
are imposed at the contracting inner grid boundary as described in 
Sect.~\ref{sec:grid}. This prescription 
torpedoes any finally conclusive analysis of neutrino-induced NS
kicks on the basis of the present models. 
We will return to this question again in the discussion of
Sect.~\ref{sec:discussion}.

\begin{table*}
\setlength{\tabcolsep}{5pt}
\centering
{
\setlength{\extrarowheight}{3pt}
\caption{Final values of explosion and NS properties for all 2D and 3D models (see Sect.~\ref{sec:auk} for definitions)}
\begin{tabular}{lccccccccc}
\hline
\hline
\multirow{2}{*}{Model}  & $E_\mathrm{exp}$ & $t_\mathrm{exp}$ & $M_\mathrm{ns}$  & $\langle v_\mathrm{ns}\rangle$  & $\langle a_\mathrm{ns}\rangle$ & $M_\mathrm{ej}$  & $P_\mathrm{ej}$  & $\langle \alpha_\mathrm{ej}\rangle$ \\
                 & $[10^{49}\,\mathrm{erg}]$ & $[\mathrm{ms}]$           & $[\msun]$        & $[\kmpers]$      & $[\mathrm{km\,s}^{-2}]$  & $[10^{-2}\,\msun]$        & $[10^{40}\mathrm{cm\,g\,s}^{-1}]$ & $[\%]$   \\
\hline
O3-dl            & 3.60             & 106              & 1.364            & $-$1.4           & 1.6              & 1.24             & 3.26             & 1.2              \\
O3-vh            & 3.78             & 102              & 1.364            & $-$0.1           & $-$0.4           & 1.23             & 3.31             & 0.1              \\
O3-vi            & 3.68             & 102              & 1.364            & $-$2.2           & $-$0.4           & 1.25             & 3.26             & 1.8              \\
O3-vl            & 2.69             & 132              & 1.365            & $-$0.0           & $-$0.3           & 1.17             & 2.68             & 0.0              \\
O3-vl2           & 2.73             & 130              & 1.365            & 0.8              & $-$1.2           & 1.15             & 2.68             & 0.8              \\
\hline
O5-dl            & 4.86             & 110              & 1.363            & $-$0.1           & 0.2              & 1.33             & 4.35             & 0.1              \\
O5-vh            & 5.70             & 96               & 1.363            & 2.1              & 0.7              & 1.32             & 4.69             & 1.2              \\
O5-vi            & 5.90             & 98               & 1.363            & 5.2              & 1.2              & 1.39             & 4.91             & 2.9              \\
O5-vl            & 4.84             & 110              & 1.364            & 2.9              & 1.5              & 1.26             & 4.18             & 1.9              \\
O5-vl2           & 5.12             & 106              & 1.364            & 2.7              & $-$0.5           & 1.25             & 4.31             & 1.7              \\
O5-vl3           & 5.13             & 116              & 1.363            & $-$0.1           & $-$2.0           & 1.36             & 4.52             & 0.1              \\
O5-vl4           & 5.23             & 108              & 1.364            & 5.8              & 0.5              & 1.27             & 4.38             & 3.6              \\
O5-vl5           & 4.85             & 110              & 1.364            & 3.2              & $-$1.5           & 1.27             & 4.19             & 2.0              \\
O5-vl6           & 4.96             & 108              & 1.364            & $-$0.7           & 0.4              & 1.23             & 4.19             & 0.5              \\
O5-vl7           & 5.18             & 106              & 1.364            & $-$0.7           & $-$1.4           & 1.31             & 4.45             & 0.4              \\
O5-vi2           & 5.67             & 98               & 1.363            & 1.3              & $-$1.1           & 1.39             & 4.78             & 0.7              \\
O5-vi3           & 5.90             & 98               & 1.362            & 1.0              & 0.2              & 1.44             & 4.99             & 0.5              \\
O5-vi4           & 5.79             & 98               & 1.363            & 0.8              & 0.8              & 1.36             & 4.80             & 0.5              \\
O5-vi5           & 5.86             & 100              & 1.363            & 0.4              & 1.7              & 1.41             & 4.89             & 0.2              \\
O5-vi6           & 6.03             & 98               & 1.362            & $-$1.9           & $-$0.1           & 1.46             & 5.07             & 1.0              \\
O5-vh2           & 5.71             & 98               & 1.363            & 0.4              & 0.8              & 1.33             & 4.74             & 0.2              \\
O5-vh3           & 6.53             & 98               & 1.361            & 0.3              & 0.2              & 1.54             & 5.49             & 0.1              \\
O5-vh4           & 5.49             & 98               & 1.363            & 0.9              & 0.9              & 1.33             & 4.61             & 0.5              \\
O5-vh5           & 6.17             & 96               & 1.362            & 1.2              & 0.1              & 1.44             & 5.09             & 0.7              \\
O5-vh6           & 6.07             & 98               & 1.363            & 1.3              & 0.3              & 1.40             & 5.00             & 0.7              \\
\hline
O8-dl            & 9.01             & 98               & 1.359            & $-$4.9           & $-$1.0           & 1.75             & 7.09             & 1.9              \\
O8-vh            & 8.96             & 94               & 1.359            & $-$2.0           & $-$0.8           & 1.77             & 7.07             & 0.8              \\
O8-vi            & 8.89             & 96               & 1.359            & 4.0              & 1.6              & 1.74             & 6.97             & 1.5              \\
O8-vl            & 8.36             & 98               & 1.360            & $-$0.6           & 0.2              & 1.72             & 6.69             & 0.3              \\
O8-vl2           & 8.52             & 100              & 1.360            & 2.4              & 1.1              & 1.70             & 6.77             & 1.0              \\
\hline
O12-dl           & 12.16            & 96               & 1.356            & $-$2.1           & $-$0.2           & 2.05             & 9.05             & 0.6              \\
O12-vh           & 12.64            & 92               & 1.356            & $-$3.7           & $-$0.3           & 2.07             & 9.25             & 1.1              \\
O12-vi           & 12.38            & 94               & 1.356            & $-$1.5           & $-$0.8           & 2.02             & 9.10             & 0.4              \\
O12-vl           & 12.25            & 96               & 1.356            & $-$0.4           & $-$0.1           & 2.05             & 9.09             & 0.1              \\
O12-vl2          & 12.23            & 96               & 1.356            & $-$2.5           & $-$0.2           & 2.04             & 9.04             & 0.7              \\
\hline
O16-dl           & 16.55            & 92               & 1.353            & 2.7              & 0.1              & 2.41             & 11.53            & 0.6              \\
O16-vh           & 16.85            & 90               & 1.352            & 0.6              & 0.0              & 2.41             & 11.59            & 0.1              \\
O16-vi           & 16.92            & 88               & 1.353            & $-$3.0           & $-$0.2           & 2.37             & 11.51            & 0.7              \\
O16-vl           & 16.42            & 92               & 1.353            & 6.3              & 0.2              & 2.40             & 11.45            & 1.5              \\
O16-vl2          & 16.59            & 92               & 1.353            & $-$3.4           & $-$0.1           & 2.39             & 11.48            & 0.8              \\
\hline
O3-3D             & 3.41       & 111            &  1.364       &  0.5          &  0.0         & 1.24     & 3.19           & 0.4              \\
O5-3D             & 5.73       & 105            &  1.362       &  0.9          &  1.4         & 1.48     & 4.91           & 0.5              \\
O8-3D             & 8.35       & 108            &  1.359       &  2.5          &  1.2         & 1.75     & 6.74           & 1.0              \\
O12-3D            & 12.00      & 107            &  1.356       &  0.4          &  $-$0.4      & 2.08     & 9.04           & 0.1              \\
O16-3D$^\ast$     & 15.92      & 101            &  1.353       &  0.3          &  $-$0.3      & 2.38     & 11.04          & 0.1              \\
\hline
\multicolumn{9}{l}{\footnotesize{$^\ast$ Simulation terminated after $0.75\,$s physical time, all other models were evolved for 1\,s.}}\\

\end{tabular}}

\label{tab:kick}
\end{table*}

%
\begin{figure}[!]
\centering
\includegraphics[width=1.\hsize]{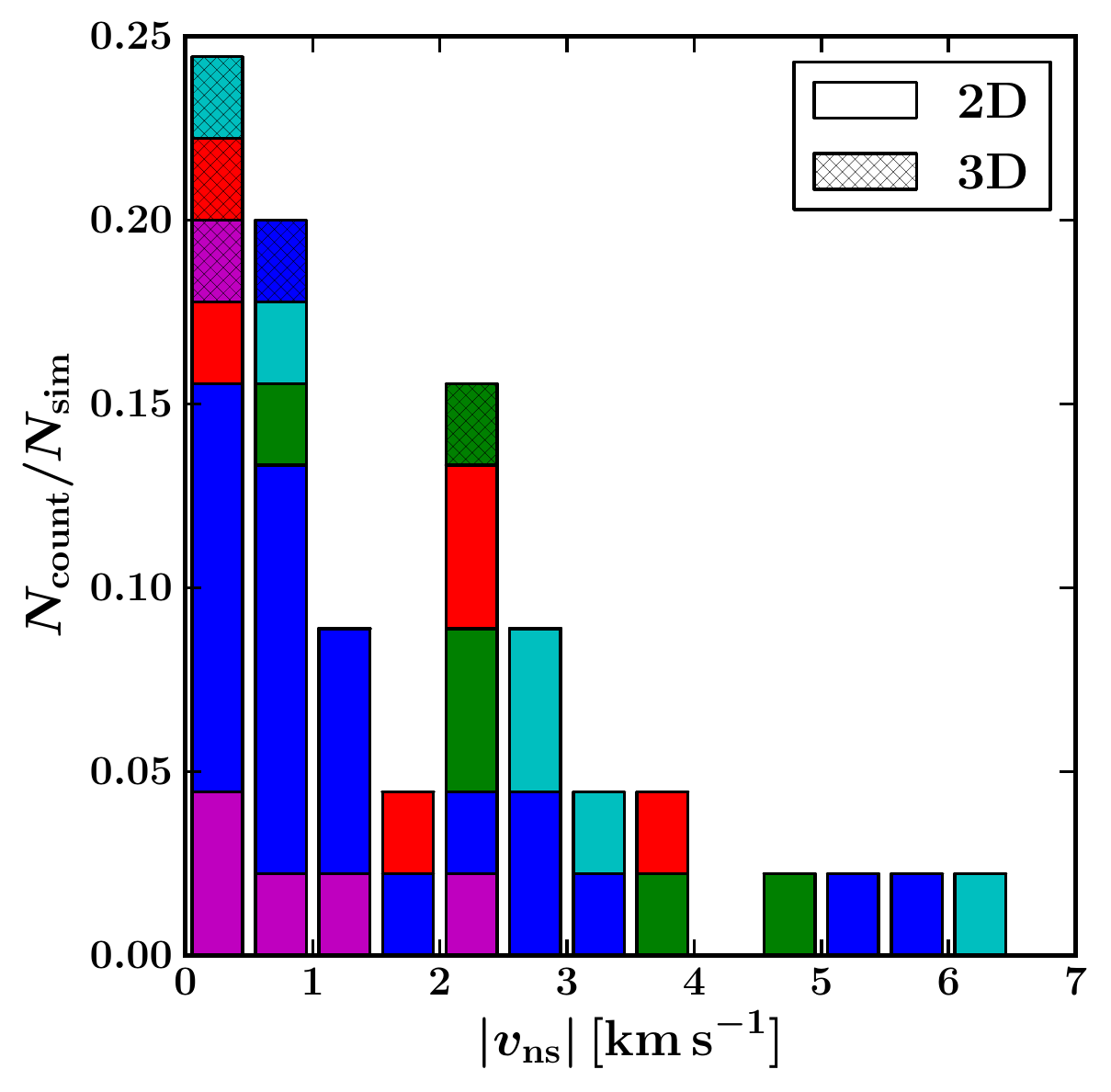}
\caption{Distribution of NS kick velocities as obtained in our
simulations, displayed by a normalized histogram (number of cases,
$N_\mathrm{count}$,
divided by total number of simulations, $N_\mathrm{sim}$) with bins of
0.5\,km\,s$^{-1}$. The colors correspond to the color
scheme introduced for our different model sets in Fig.~\ref{fig:eexp}:
\fcolorbox{black}{pymagenta}{\rule{0pt}{6pt}\rule{6pt}{0pt}} O3, 
\fcolorbox{black}{pyblue}{\rule{0pt}{6pt}\rule{6pt}{0pt}} O5, 
\fcolorbox{black}{pygreen}{\rule{0pt}{6pt}\rule{6pt}{0pt}} O8, 
\fcolorbox{black}{pyred}{\rule{0pt}{6pt}\rule{6pt}{0pt}} O12, and 
\fcolorbox{black}{pycyan}{\rule{0pt}{6pt}\rule{6pt}{0pt}} O16.
Note that the histogram is
just a visual representation of our results for the different sets
of parameter values but does not correspond to a probability distribution
for NS kick velocities expected in Nature.}
\label{fig:hist}
\end{figure}

\begin{figure*}[!]
\centering
\includegraphics[width=0.9\textwidth]{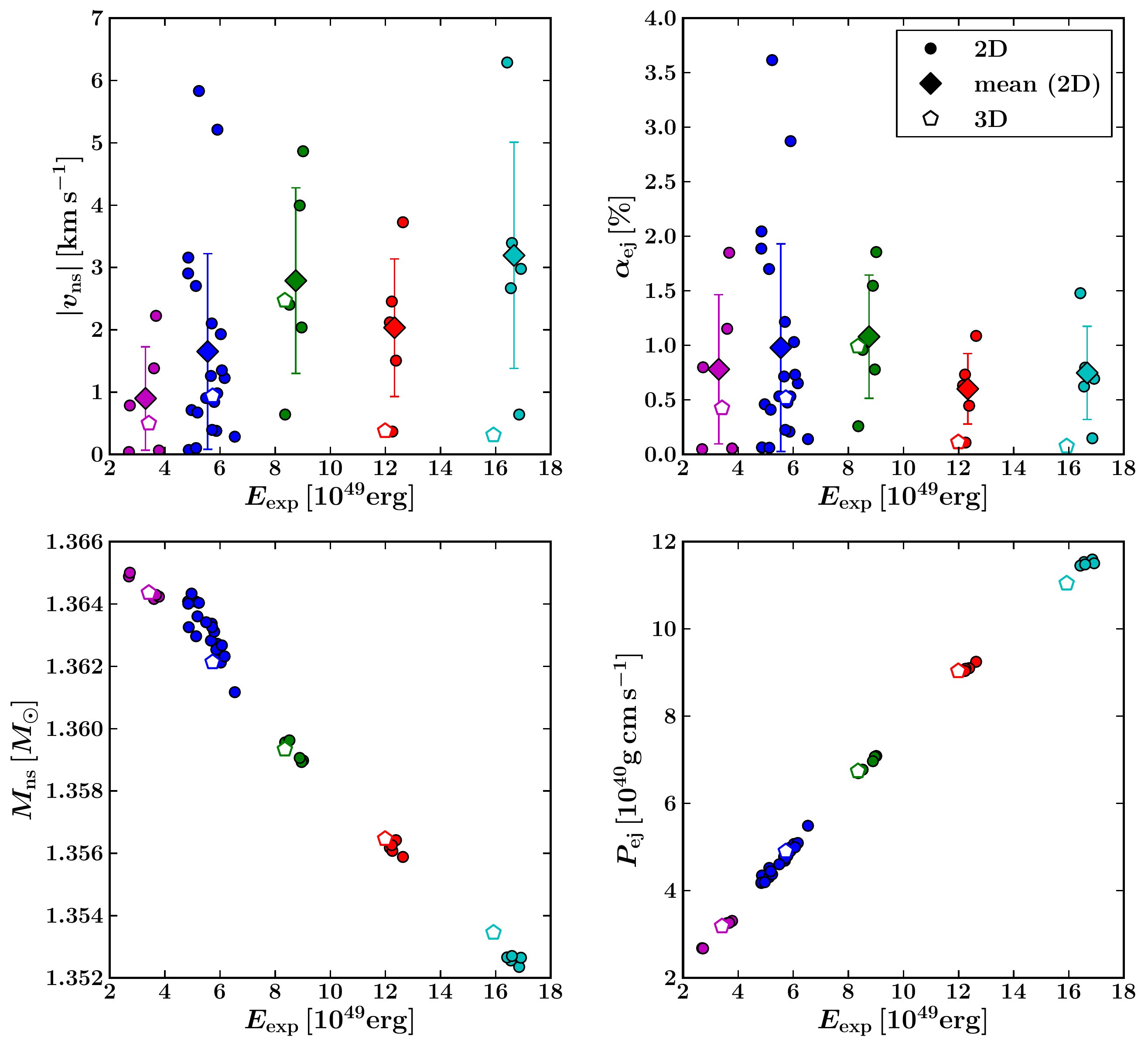}
\caption{Dependences of the final values of NS kick velocity (\textit{upper left}),
ejecta-momentum asymmetry parameter (\textit{upper right}), NS mass (\textit{lower left}),
and radial ejecta momentum (\textit{lower right}) on the SN explosion energy.
Filled colored circles represent data from the 2D models according to the 
color scheme introduced in Fig.~\protect\ref{fig:eexp} for our five model sets.
Filled colored diamonds mark the mean values of all 2D
models belonging to the different model sets, and vertical bars indicate the 
corresponding standard deviations. Open pentagons stand for the results of our
3D models. All 3D models show the tendency to kick velocities and ejecta-momentum
asymmetries below the average of the 2D cases.}
\label{fig:stat}
\end{figure*}

\subsubsection{Compilation of simulation results}
\label{sec:compilation}
We evaluate the NS kick for all models according to Eq.~(\ref{eq:momcon}).
The corresponding results are listed in Table~\ref{tab:kick}, and Figure~\ref{fig:hist}
depicts a histogram, normalized w.r.t.\ the total number of simulation runs,
for the absolute values of the NS kick velocities of all computed models, 
binned in intervals of 0.5\,km\,s$^{-1}$.
None of the computed cases yields a kick larger than 7\,km\,s$^{-1}$ after 
one second of simulated evolution, and in more than half of the model runs the NS
kick is less than 1.5\,km\,s$^{-1}$.
Especially all of the 3D models tend to lie on the low-velocity side of the distribution
and the highest kick in 3D (model O8-3D) is less than half the maximum value obtained
for the 2D models. 

It is clear from these results that ECSNe are unlikely to produce hydrodynamical NS kicks
of more than a few kilometers per second. This is in stark contrast to explosions of 
massive Fe-core progenitors, where explosion models yield average kick velocities of
several hundred km\,s$^{-1}$ by the same acceleration mechanism
\citep{Schecketal2004,Schecketal2006,Wongwathanaratetal2010,Wongwathanaratetal2013,
Nordhausetal2010,Nordhausetal2012,Bruennetal2016,Muelleretal2017},
in agreement with observations of many radio pulsars as well as NSs in SN remnants.
The NS kicks we obtain for ECSNe are even an order of magnitude lower than the kick
velocities (up to some 10\,km\,s$^{-1}$) found for explosions of ultrastripped SN progenitors
by \citet{Suwaetal2015,Muelleretal2018}.

From our discussion of the explosion dynamics of ECSNe (Sect.~\ref{sec:ex_prop})
and of the basic physics of the gravitational tug-boat mechanism providing
the NS acceleration (Sect.~\ref{sec:kickphysics}), we can understand the reasons 
for the inefficiency of ECSNe in kicking the NS.
Because of the extremely steep density decline at the edge of the
degenerate core, the SN shock runs outward with very high speed, allowing
also the rapid expansion of neutrino-heated matter in its wake. Although
convective overturn develops above the gain radius, the high-entropy
plasma expands so quickly that the buoyant Rayleigh-Taylor plumes freeze 
out in a short-wavelength pattern. This implies that a sizable dipole mode is
absent and the momentum asymmetry therefore remains small.
Moreover, efficient momentum transfer from the anisotropic
ejecta to the NS is prevented by the fast outward acceleration and by extremely low
densities in the surroundings of the PNS, entailing that very little mass carries the
asymmetries at the beginning of the explosion.

In SN explosions of massive iron-core progenitors significant dipolar asymmetry of
the neutrino-heated ejecta is a generic feature 
\citep[see, e.g.,][]{Wongwathanaratetal2013,Lentzetal2015,
Melsonetal2015a,Muelleretal2017,Ottetal2018}. It originates from the development
of low-mode mass motions associated with convective or SASI activity in the volume
between PNS and stalled SN shock. 
These instability modes take about 100\,ms to reach
the nonlinear stage, and SASI growth is particularly fast during phases of shock
stagnation or even retraction. In ECSNe, however, the extremely steep density drop 
outside of the O-Ne-Mg core (or, equivalently, the extremely small compactness value
of the core\footnote{The 8.8\,$M_\odot$ O-Ne-Mg core progenitor employed in our 
study has compactness values (before collapse as well as at core 
bounce) of $\xi_{1.4} = 1.1\times 10^{-5}$ and
$\xi_{1.5} = 6.6\times 10^{-6}$, which correspond to radii of
more than $10^{13}$\,cm in this star. This demonstrates the extremely dilute 
environment of the degenerate core of $\sim$1.36\,$M_\odot$. The quoted numbers
for $\xi_{1.4}$ and $\xi_{1.5}$ are consistent with those of a modern O-Ne-Mg-core
progenitor provided by Alexey Tolstov and Ken Nomoto (private communication, 2017).
Although the density profiles of the old and new progenitors differ in the immediate
surroundings of the degenerate core, these regions are so tenuous in both cases
(with maximum densities of the H/He envelope around 10\,g\,cm$^{-3}$\,!)
that the explosion dynamics of the ECSNe are the same (work in progress).
The compactness values of these O-Ne-Mg-core progenitors are therefore much lower
than the value of $\xi_{1.5} = 2.3\times 10^{-4}$ of the zero-metallicity
9.6\,$M_\odot$ iron-core progenitor considered in \citet{Sukhboldetal2016},
which in turn possesses a considerably steeper density 
decline exterior to the iron core than any of the solar-metallicity
low-mass pre-SN models computed by \citet{WoosleyHeger2015} 
\citep[see figure~3 in][]{Sukhboldetal2016}.})
permits continuous shock expansion (even during the first 100\,ms
after bounce) without any phase of shock stagnation 
\citep[see][]{Jankaetal2008}. This disfavors the growth of dipolar asymmetry modes
in the neutrino-heated layer as visible in Fig.~\ref{fig:elena_plots}, where the
Rayleigh-Taylor plumes reflect a high-order spherical harmonics pattern which
inflates very quickly and basically self-similarly already after the first 
$\sim$100\,ms.

%
\subsubsection{Ejecta anisotropy and kick systematics}
\label{sec:auk}
Besides listing the absolute values of the final NS velocity, $v_\mathrm{ns}$,
and acceleration, $a_\mathrm{ns}$, at the end of our simulations, 
Table~\ref{tab:kick} also compiles quantities that are diagnostically 
relevant for the SN explosion, i.e., the explosion energy, $E_\mathrm{exp}$;
the time for the onset of the explosion, $t_\mathrm{exp}$, which is defined
as the instant when the (diagnostic) explosion energy exceeds $10^{49}$\,erg
(the reader is reminded here that the corresponding post-bounce time is 
$t_\mathrm{exp}+18$\,ms); the baryonic mass of the NS, $M_\mathrm{ns}$; the
ejecta mass, $M_\mathrm{ej}$, defined as the gas mass enclosed in the shell between
the NS on the one side and the SN shock on the other. 
All of these quantities (except for $t_\mathrm{exp}$)
are given at the end of the simulations, which is after 1\,s of computed
evolution except for model O16-3D, which terminated after 750\,ms.
The values of $v_\mathrm{ns}$ result from the requirement of momentum conservation
of ejecta gas plus NS in the hydrodynamical simulation (Eq.~\ref{eq:momcon});
the corresponding time derivative yields $a_\mathrm{ns}$.
Note that the notation $\langle X\rangle$ indicates that the tabulated value
of quantity $X$ is the average over the last 20\,ms of the respective
simulation. 

The last two columns of Table~\ref{tab:kick} provide values for the quantities 
$P_\mathrm{ej}$ and $\alpha_\mathrm{ej}$, which are the (radial) ejecta momentum
and the momentum-asymmetry parameter of the ejecta, respectively. These two quantities
allow for a better understanding of the origin of systematic trends in the obtained 
NS kicks. They are defined as \citep[following][]{Schecketal2006}
\begin{equation}
\alpha_\mathrm{ej} = \frac{\left| \vect{P}_\mathrm{gas}\right|}{P_\mathrm{ej}}\,,
\label{eq:alphaej}
\end{equation}
which relates the net gas momentum $\left| \vect{P}_\mathrm{gas}\right|$ (see
Eq.~\ref{eq:momcon} and text below this equation) to the scalar quantity
\begin{equation}
P_\mathrm{ej} = \int_{R_\mathrm{ns}}^{R_\mathrm{s}} 
\rho\left|\vect{v}\right|\,\diff V\,.
\label{eq:pej}
\end{equation}
According to Eq.~(\ref{eq:pej}), $P_\mathrm{ej}$ represents the total (basically 
radial, because $|\vect{v}| = v_r$ to high accuracy) momentum available in the 
volume between NS surface (at $R_\mathrm{ns}(\theta,\phi)$) and SN shock (at radius 
$R_\mathrm{s}(\theta,\phi)$). Using these definitions, we can write the magnitude
of the NS kick velocity in terms of the radial ejecta momentum and the 
momentum-asymmetry parameter as
\begin{equation}
\left|\vect{v}_\mathrm{ns}\right|=\alpha_\mathrm{ej}\,\frac{P_\mathrm{ej}}{M_\mathrm{ns}}\,.
\label{eq:vns}
\end{equation}

The parameter $\alpha_\mathrm{ej}$ (which can vary between 0 and 1) suits to 
characterize the ejecta morphology. Under the assumption that the explosion energy
is dominated by the kinetic energy of the expanding (anisotropic) 
ejecta\footnote{The validity of this assumption for the considered
explosion models of ECSNe can be concluded from a closer inspection of 
Figs.~\ref{fig:2dsto}--\ref{fig:3dsto_2}. In neutrino-driven explosions of ECSNe
the explosion energy is naturally carried by the neutrino-heated, anisotropic,
high-entropy matter ejected from the near-surface layers of the degenerate core,
the bulk of whose mass ends up in the newly formed NS. This ejected matter is 
visualized by the high-entropy plumes and their surrounding material in 
Figs.~\ref{fig:2dsto}--\ref{fig:3dsto_2}. At $t\gtrsim 500$\,ms after core bounce,
these ejecta expand basically self-similarly with a velocity of about 
$v_\mathrm{ej}\sim 15,000$--30,000\,km\,s$^{-1}$, as can be seen in the 
displayed images. On the other
hand, for entropies of $\sim$15--20\,$k_\mathrm{B}$ per nucleon in the
plumes and considerably lower values in their surroundings, and temperatures
$T\lesssim 10^9$\,K, the sound speed, $c_\mathrm{s}$,
is several 1,000\,km\,s$^{-1}$ at most. Since $v_\mathrm{ej}\gg c_\mathrm{s}$,
kinetic energy dominates the interal energy of this matter.}
(or just a constant fraction of it), i.e.\
$E_\mathrm{exp}\propto E_\mathrm{kin}$, $P_\mathrm{ej}$ is related to the 
explosion energy by $E_\mathrm{exp}\propto\frac{1}{2} P_\mathrm{ej}^2 M_\mathrm{ej}^{-1}$.
This implies
\begin{equation}
P_\mathrm{ej} \propto \sqrt{2\,M_\mathrm{ej}\,E_\mathrm{exp}} \,.
\label{eq:pme}
\end{equation}
The excellent validity of this relation for ECSNe can be concluded from the 
discussion following below and the results shown in Fig.~\ref{fig:stat}.
For more massive Fe-core SN explosions it will be demonstrated in a
forthcoming paper \citep{WongwathanaratJanka2018}.

In neutrino-driven explosions, neutrino heating lifts matter from the vicinity
of the PNS to a gravitationally less or even marginally bound state (i.e., to 
specific binding energy closer to zero), while subsequent energy release by the
recombination of free nucleons to heavy nuclei (up to $\sim$8.8\,MeV per nucleon)
provides the excess energy for the explosion
\citep[see][]{Janka2001,MarekJanka2009,Mueller2015}.\footnote{In explosions of
massive iron-core progenitors, the net energy per nucleon obtained by neutrino
heating and nucleon recombination in the ejecta is typically 5--7\,MeV
\citep{Janka2001,MarekJanka2009,Mueller2015}. In the considered ECSNe we obtain
a value of $\sim$6\,MeV per nucleon, which can be concluded from the data in 
the lower left panel of Fig.~\ref{fig:stat}.}
Therefore one expects that the explosion energy grows with the mass of the
neutrino-heated ejecta,
\begin{equation}
E_\mathrm{exp} \propto M_\mathrm{ej} \sim M_\mathrm{s} - M_\mathrm{ns} \,,
\label{eq:eexpmej}
\end{equation}
where $M_\mathrm{s}$ is the mass enclosed by the SN shock during the first
second(s) of the explosion. For the considered ECSN simulations, this expectation 
is confirmed by Fig.~\ref{fig:stat} (lower left panel), which shows that 
$E_\mathrm{exp}$ increases linearly with decreasing NS mass ${M_\mathrm{ns}}$. 
As more matter is expelled in the explosion by neutrino heating instead of being
integrated into the NS, the explosion energy increases by the same degree.
The extra mass expelled enlarges the mass available between NS and shock,
$M_\mathrm{s} - M_\mathrm{ns}$.

Using the proportionality relation of Eq.~(\ref{eq:eexpmej}) in
Eq.~(\ref{eq:pme}) yields
\begin{equation}
P_\mathrm{ej} \propto E_\mathrm{exp}\,.
\label{eq:pexp}
\end{equation} 
This relation is fully in line with Fig.~\ref{fig:stat}, lower right panel. 
Equation~(\ref{eq:pexp}) in Eq.~(\ref{eq:vns}) finally leads to
\begin{equation}
\left|\vect{v}_\mathrm{ns}\right|\propto 
\frac{\alpha_\mathrm{ej}}{M_\mathrm{ns}}\,E_\mathrm{exp} 
\label{eq:vns2}
\end{equation}
\citep[see also][equation~11 there]{Janka2017}.

Our result of Eq.~(\ref{eq:vns2}) explains what we see for our large
sample of 2D models in the upper left panel of Fig.~\ref{fig:stat}:
Stochastic variations of $|\vect{v}_\mathrm{ns}|$ stem from such
variations in the asymmetry parameter $\alpha_\mathrm{ej}$ 
(Fig.~\ref{fig:stat}, upper right panel), on top of which a linear
trend of increase of $|\vect{v}_\mathrm{ns}|$ with $E_\mathrm{exp}$
is superimposed. Because of the tendency of higher kicks for higher
explosion energies, also the variance of the kicks exhibits this
trend. 

For all five energy sets the kick velocity of the 3D model is below 
the average of the 2D models. This can be expected from the axisymmetric
geometry of 2D simulations, which favors the formation of downflows at 
the poles (see Fig.~\ref{fig:2dsto}). In that sense O8-3D is peculiar.
Because of the stochastic development of ejecta
asymmetries in the nonlinear stage of the growth of hydrodynamic
instabilities, model O8-3D accidentally is
the only 3D case that features a similarly prominent 
downflow at the end of the simulation as many of the 2D models.
For this reason model O8-3D is the only 3D case with a NS kick
close to the mean value of the 2D runs.

The values of the asymmetry parameter $\alpha_\mathrm{ej}$ 
(Fig.~\ref{fig:stat}, upper right panel) are below $\sim$3.5\%,
on average around 1\%. This is a factor of 10 smaller than
those in typical iron-core SN explosions, where values up to roughly
$\sim$35\% were found with an average around 10\% 
\citep[][figure~11 there]{Schecketal2006}.

An interesting question concerns a possible dependence of the values
of the momentum-asymmetry parameter 
$\alpha_\mathrm{ej}$ on the explosion energy.
The asymmetries of the 2D models exhibit a (slight) indication of a
maximum at intermediate explosion energies. The 3D models seem to 
mirror such a trend, but this has to be taken with caution in
view of the fact that we have only one 3D simulation for each 
value of the explosion energy. 
Moderate $E_{\mathrm{exp}}$ therefore might favor the highest 
anisotropies. Such a local peak could be imagined as a consequence
of two competing effects:
Weak neutrino heating for cases with low explosion energies
hampers the growth of large asymmetries by strong convective activity.
In contrast, high
explosion energies due to strong neutrino-energy deposition imply 
very fast expansion of the convectively perturbed, neutrino-heated
ejecta. Therefore the matter in the close vicinity of the NS is 
quickly replaced by the spherically symmetric neutrino-driven wind,
which does not contribute to the recoil acceleration of the NS.
Hence, O3 models (i.e., models with the lowest explosion energies)
yield NS kicks $|\vect{v}_\mathrm{ns}|$ 
that are small on average because of low values of $\alpha_\mathrm{ej}$
with little variance as well as low available momentum $P_\mathrm{ej}$.
Likewise, O12 and O16 models (with their highest explosion energies
in our sample) feature relatively small $\alpha_\mathrm{ej}$,
but due to the higher values of $P_\mathrm{ej}$ their average NS kicks
are on the larger side, and also the variance of the kick velocities
is considerable (upper left panel of Fig.~\ref{fig:stat}). 
This means that increasing ejecta momentum compensates for the
decrease of the anisotropy parameter to produce relatively large kicks.
Of course, finally conclusive results for the dependence of 
$\alpha_\mathrm{ej}$ on $E_{\mathrm{exp}}$ would require still more
models, in particular an extensive sample of 3D calculations.

\section{Discussion of Implications for Crab}
\label{sec:discussion}

What caused the kick of the Crab pulsar? In view of our results 
presented in Sect.~\ref{sec:kickresults} the answer is far from
being clear. If Crab is the remnant of an O-Ne-Mg core progenitor,
a hydrodynamical kick mechanism for its pulsar is strongly 
disfavored by the extremely low NS recoil velocities obtained in our
simulations.

A possible alternative might be a postnatal acceleration of 
the pulsar due to electromagnetic radiation produced by an
off-centered, rotating magnetic dipole \citep{HarrisonTademaru1975}.
The NS kick by this electromagnetic rocket effect is attained
on the spin-down timescale and is imparted along the spin axis
of the pulsar. The NS obtains its proper motion at the expense 
of kinetic energy associated with its rotation.
This scenario appears attractive because of the impressive 
radiation produced by the Crab pulsar even at the present age 
and because of the observed rough spin-kick alignment of this
compact object, provided this alignment is not purely incidental
\citep[for a discussion with references, see][]{Laietal2001}.

However, a quantitative analysis reveals that this explanation
is unlikely. Pushing all involved parameters to their extreme
limits, a pulsar kick of at most 
\begin{equation}
v_\mathrm{HT,kick}^\mathrm{max} \approx 5.05\,\, R_{12}^2\,
\left(\frac{20\,\mathrm{ms}}{T_\mathrm{ns,i}}\right)^{\! 2}
\,\mathrm{\frac{km}{s}}
\label{eq:HTkick}
\end{equation}
can be expected by the Harrison-Tademaru effect
\citep[][equation~5 there with pulsar spin frequency
$\nu = T_\mathrm{ns}^{-1}$ and $\nu \ll \nu_\mathrm{i}$]{Laietal2001}.
In Eq.~(\ref{eq:HTkick}), $R_{12} = R_\mathrm{ns}/(12\,\mathrm{km})$
is the NS radius and $T_\mathrm{ns,i}$ is the initial NS spin period. 
This means that in order to achieve a recoil of 160\,km\,s$^{-1}$ 
by the emission of electromagnetic energy, an initial value of
$T_\mathrm{ns,i}\sim 3.5$\,ms is required, in conflict with the
estimated $\sim$19\,ms birth period of the Crab pulsar 
\citep{BejgerHaensel2003,ManchesterTaylor1977}. 
Moreover, the output of spin-down 
energy of a 3.5\,ms pulsar, $\Delta E_\mathrm{rot}
\sim \frac{1}{2}I_\mathrm{ns}(2\pi/T_\mathrm{ns,i})^2\approx
\frac{1}{2}\left(\frac{2}{5}M_\mathrm{ns}R_\mathrm{ns}^2\right)
(2\pi/T_\mathrm{ns,i})^2 
\sim 2.8\times 10^{51}M_{1.5}R_{12}^2
(T_\mathrm{ns,i}/3.5\,\mathrm{ms})^{-2}$\,erg (with the NS mass
$M_{1.5} = M_\mathrm{ns}/(1.5\,M_\odot)$ and NS moment of inertia
$I_\mathrm{ns}$), is incompatible with the energetics and dynamics
of the Crab Nebula (\citealt{YangChevalier2015}; but see 
\citealt{Atoyan1999}
for a different opinion, arguing that basically all of the 
injected energy could have been radiated away). 

A second alternative possibility for explaining the proper motion of
the Crab pulsar could be the SN-induced breakup of a close binary
system, in which the Crab progenitor would have been one
component. This Blaauw effect \citep{Blaauw1961} could account for the
inferred $\sim$160\,km\,s$^{-1}$ of the NS, but evidence for
a former companion star does not exist. Moreover, if this scenario
were true, the current interpretation of the observed 
southeast-northwest asymmetry of the Crab's synchrotron
nebula as a consequence of the proper motion of the pulsar would
require revision. A displacement of the pulsar
from the explosion center \citep[][figure~7 therein]{Hester2008}
would not be compatible with a binary disruption by the Crab SN,
in which case the NS and the center of the expanding ejecta cloud
would move with the same speed of the progenitor's previous orbital 
motion in the binary system.

A third possible alternative could be a low-mass iron-core progenitor
for the origin of Crab. SN explosions of such progenitors in the 
$\sim$9\,$M_\odot$ to $\sim$12\,$M_\odot$ range \citep{WoosleyHeger2015} 
eject similar amounts of stellar debris with similarly low explosion 
energies \citep[of order $\sim$$10^{50}$\,erg; see][]{Melsonetal2015,
Radiceetal2017}
as ECSNe, and they also underproduce $^{56}$Ni compared to core-collapse
SNe of more massive progenitors \citep{Ertletal2016,Sukhboldetal2016}. 
Nevertheless, low-mass iron-core progenitors
can possess considerably higher values for the 
core compactness (i.e., more shallow density profiles exterior to the
iron core). There is a considerable variation in the density
profiles between different representatives in this mass
range (\citealt{Sukhboldetal2016}, figure~3, and \citealt{Janka2017a},
figure~2), corresponding to an increase of $\xi_{1.5}$
from about $7.5\times 10^{-3}$ for a 9.0\,$M_\odot$
progenitor to $\xi_{1.5} \sim 0.094$ for a 9.5\,$M_\odot$
star and $\xi_{1.5}$ between 0.44 and 0.58 for models
between 10.0\,$M_\odot$ and 12.0\,$M_\odot$. Therefore such
stars might explode with sufficient ejecta momentum asymmetry
to produce NS kick velocities in the ballpark of 160\,km\,s$^{-1}$. 
This argument receives support by the results of
\citet{Suwaetal2015,Muelleretal2018}, who reported (non-converged)
kick velocities of several 10\,km\,s$^{-1}$
up to $\sim$75\,km\,s$^{-1}$ within less
than one second of post-bounce evolution for a small set of
2D neutrino-driven SN simulations of ultra-stripped SNe~Ic
progenitors (with compactness values $\xi_{1.4}$ between
$\sim$0.47 and $\sim$0.97). 
However, further investigation is needed to clarify whether low-mass
iron-core progenitors are compatible with the ejecta composition of 
the Crab Nebula, whose high helium abundance and relatively
oxygen-poor filaments served as argument
for an O-Ne-Mg-core progenitor (\citealt{Nomotoetal1982,DavidsonFesen1985};
see also the concise overview of Crab properties and corresponding
references provided by \citealt{Smith2013}).

Finally, a fourth possibility is that the Crab pulsar attained its kick
by anisotropic neutrino emission during the neutrino-cooling phase of
the hot PNS. The corresponding neutrino-induced kick velocity can be
written as
\begin{equation}
v_{\nu,\mathrm{kick}} = 167\,\frac{\alpha_\nu}{0.005}\,
\frac{E_\nu}{3\!\times\! 10^{53}\,\mathrm{erg}}\,
\left(\!\frac{M_\mathrm{ns}}{1.5\,M_\odot}\!\right)^{\! -1}\,
\mathrm{\frac{km}{s}}\,,
\label{eq:nukick}
\end{equation}
where $\alpha_\nu = P_{z,\nu}/P_\nu$ is the neutrino momentum asymmetry
parameter and $P_\nu = E_\nu/c$ is the total radial momentum associated
with a neutrino-energy loss $E_\nu$ \citep[see][]{Schecketal2006}. 
The parameter $\alpha_\nu$ is related to the amplitude $a_\mathrm{d}$
of the dipole component (oriented in $z$-direction) 
of a neutrino emission asymmetry, 
$\diff L_\nu(t)/\diff \Omega = L_\nu(t)(1+a_\mathrm{d}\cos\theta)/(4\pi)$,
by the relation
$\alpha_\nu = a_\mathrm{d}/3$. Therefore Eq.~(\ref{eq:nukick})
implies that an emission dipole of roughly 1.5\% of the monopole, applied
to the total neutrino-energy loss $E_\nu$, is sufficient to obtain a kick
of the magnitude of the Crab pulsar.

One way to generate considerable dipolar asymmetry of the
neutrino emission is by the presence of very strong magnetic fields
in the NS interior 
\citep[see, e.g.,][]{Bisnovatyi-Kogan1993,Socratesetal2005,Kusenkoetal2008,
SagertSchaffner2008,Maruyamaetal2012}, which can modify
the neutrino interactions in the dense matter. Corresponding
neutrino-induced kicks of some 100\,km\,s$^{-1}$ typically require 
dipolar field asymmetries of order $10^{15}$\,G or more. 
But this requirement is in conflict with the
fact that the Crab pulsar has an ordinary dipolar surface
field of about $4\times 10^{12}$\,G \citep[e.g.,][]{Michel1991},
and interior fields of order $10^{15}$\,G would be purely
speculative. On the other hand, neutrino-emission
dipoles of several percent amplitude have been found in association 
with the newly discovered dipolar lepton-emission self-sustained asymmetry
(LESA; \citealt{Tamborraetal2014,Jankaetal2016}; see also the recent
confirmation by \citealt{OConnorCouch2018}) in full-scale 3D simulations
of NS birth. LESA does not require the presence of a magnetic field.
The corresponding emission dipole exhibits a stable or
only slowly changing direction and can amount to a few percent
of the added luminosities of neutrinos and antineutrinos
of all species. However, current 3D simulations have been
able to follow this phenomenon in the nascent NS
only for the first 500--700\,ms after core bounce. It is
therefore unclear how long a significant dipolar asymmetry
of the neutrino emission can persist and whether it can
reach an overall amplitude of order 1--2\% of the total
(i.e., time-integrated) neutrino-energy loss, which is needed
to explain a NS kick of about 160\,km\,s$^{-1}$ (see
Eq.~\ref{eq:nukick}). If LESA is a generic phenomenon during
the neutrino-cooling phase of all hot, new-born NSs, it
would lead to a relatively high ``floor value'' for NS kicks
(see \citealt{BrayEldridge2016} and \citealt{BrayEldridge2018}
for a corresponding conjecture
based on combining measured NS proper motions and stellar
population modeling of stars evolving to SNe). However, such
a ubiquitous NS acceleration mechanism, which would act in
all cases, might be in conflict with
constraints associated with a population of low-kick NSs
in binary systems \citep{Podsiadlowskietal2004,Schwabetal2010,
BeniaminiPiran2016,Taurisetal2017,Kruckowetal2018}.

In summary, the origin of the kick of the Crab pulsar remains a puzzle
and could be a smoking gun for either the nature of the progenitor  
of the Crab SN or neutrino-related emission asymmetries of the nascent
NS.

\section{Summary and Conclusions}
\label{sec:conclusions}

We have presented results of 40 2D and five 3D simulations of
neutrino-driven ECSN explosions of an O-Ne-Mg-core progenitor,
considering models with parametrically tuned explosion energies 
between about $3\times 10^{49}$\,erg and roughly
$1.6\times 10^{50}$\,erg.
This range of energies brackets the results of previous
self-consistent core-collapse and explosion simulations 
\citep{Kitauraetal2006,Jankaetal2008,
Huedepohletal2010,Fischeretal2010,vonGroote2014,Radiceetal2017}
and the estimated explosion energy of the Crab SN
\citep{YangChevalier2015}, which is discussed as a possible
candidate for an ECSN 
\citep[e.g.,][]{Nomotoetal1982,Hillebrandt1982,Tominagaetal2013,Smith2013}.

The goal of our study was to explore the hydrodynamic NS kicks 
that are associated with anisotropic mass ejection in this type
of SN explosion. The asymmetries originate from a brief episode
of growth of Rayleigh-Taylor instability in the convectively
unstable neutrino-heated region exterior to the gain radius.
The emerging inhomogeneities in the mass and energy 
distributions of the neutrino-heated ejecta exert hydrodynamical
and gravitational forces on the NS and thus accelerate the NS
via the gravitational tug-boat mechanism. The
typical NS kick velocities that we obtained in our ECSN models
are of the order of kilometers per second.

Our results support previous theoretical arguments that NS kicks
in ECSNe should be smaller on average compared to NS kicks 
associated with SNe of more massive iron-core progenitors 
\citep{Podsiadlowskietal2004,Janka2017}. The extremely low compactness
values of the stellar core ($\xi_{1.4} = 1.1\times 10^{-5}$ and
$\xi_{1.5} = 6.6\times 10^{-6}$) enable a very rapid outward 
acceleration of the SN shock, which also allows the postshock layer
to expand rapidly in the wake of the shock. 
For this reason the merging process of the Rayleigh-Taylor 
plumes of neutrino-heated, buoyant plasma to bigger structures
freezes out in a higher-order spherical harmonics pattern with
insignificant amplitudes of dipolar or quadupolar deformation
modes.

This ECSN-typical dynamical scenario has three consequences that
conspire to disfavor large NS kicks: First, the dipolar momentum 
asymmetry (measured by the parameter $\alpha_\mathrm{ej}$;
Eq.~\ref{eq:alphaej})
is only of order one percent and thus about 10 times smaller than 
for typical iron-core SNe. Second, the low explosion energy of
ECSNe (around $10^{50}$\,erg) implies a low radial momentum of the
ejecta and also reduces the NS kick (see Eqs.~\ref{eq:pexp} and
\ref{eq:vns2}). Third, the extremely fast expansion of the convectively
perturbed ejecta diminishes the time over which the anisotropic
forces responsible for the gravitational tug-boat acceleration can
act on the NS. These effects explain why our simulations yield 
NS kick velocities of only a few kilometers per second at most, 
which is a factor of 100 below those typical of iron-core SNe 
\citep[see][]{Schecketal2006,Wongwathanaratetal2013}. 
Our extensive set of 40 2D and five 3D simulations for a relevant
range of explosion energies accounts for stochastic variations of
the ejecta asymmetry and thus renders this finding a statistically
consolidated result. In view of our pool of results,
testing different explosion energies and initial perturbations, it 
appears very unlikely that current ECSN models can 
explain NS kicks of more than 100\,km\,s$^{-1}$ with any significant
statistical probability.

The low-velocity NS kicks that we obtained in our ECSN models are
compatible with the possibility of a dichotomous distribution of NS 
kicks \citep[e.g.,][and references therein]{Katz1975,Katz1983,
Podsiadlowskietal2004,Schwabetal2010,BeniaminiPiran2016},
and they also back up the possibility of a large
population of double compact objects as gravitational-wave sources
\citep[e.g.,][]{Taurisetal2017,Kruckowetal2018}. However, such low 
NS kicks are in tension with the hypothetical explanation
of the Crab Nebula as the gaseous remnant of an ECSN 
\citep[e.g.,][]{Nomotoetal1982,Hillebrandt1982,Tominagaetal2013,Smith2013}, 
because the Crab pulsar's measured proper motion of 
$\sim$160\,km\,s$^{-1}$ is two orders of magnitude higher than the
average NS kicks found in our ECSN explosion models. Such 
velocities are out of reach by hydrodynamical kicks in ECSNe
even for statistical outliers.

A postnatal acceleration of the Crab pulsar by the electromagnetic 
rocket effect \citep{HarrisonTademaru1975} might appear attractive 
because of the apparent aligment of the pulsar's spin and kick.
However, this explanation is strongly disfavored by its 
requirement of a short initial spin period of less than 4\,ms
(see detailed discussion in Sect.~\ref{sec:discussion}), whose
associated release of electromagnetic power is more than an order
of magnitude higher than limits set by a detailed energetic 
analysis of the Crab Nebula \citep{Hester2008,YangChevalier2015}.

Alternatively, attributing the proper motion of the Crab pulsar 
to the disruption of a binary system in the event of SN~1054 is
in tension with the current picture of the internal structure and
dynamics of the Crab Nebula. If the NS's space velocity had
originated from the disruption of a close binary system, the
pulsar would not move relative to the mass-center of the ejecta 
gas, and a displacement of the pulsar out of the center of the
explosion \citep[][figure~7]{Hester2008} would not apply.
Moreover, the spin-kick alignment of the Crab pulsar would imply
that the spin axis was close to the orbital plane, which is not
the most plausible configuration of spin and orbital angular 
momentum.

Two other possible explanations of the Crab pulsar's proper 
motion appear more likely. Either the progenitor of SN~1054 was
a low-mass star with an iron core instead of the hypothesized
O-Ne-Mg-core progenitor. Or the Crab pulsar received its kick
velocity by anisotropic neutrino emission rather than being
kicked by the hydrodynamical mechanism associated with 
asymmetric mass ejection in the SN explosion.

Neutrino-induced NS kicks of $\gtrsim$100\,km\,s$^{-1}$
may be a consequence of anisotropic
neutrino emission associated with an ultra-strong dipole
magnetic field of about $10^{15}$\,G or more 
\citep[e.g.,][]{Bisnovatyi-Kogan1993,Socratesetal2005,Kusenkoetal2008,
SagertSchaffner2008,Maruyamaetal2012}.
Since the Crab pulsar is not a magnetar,
the existence of such enormous magnetic fields in its
interior would be pure speculation.
A dipolar asymmetry of the neutrino emission
is also associated with the LESA (lepton-emission
self-sustained asymmetry) instability newly discovered in 3D SN
simulations 
\citep{Tamborraetal2014,Tamborraetal2014a,Jankaetal2016,OConnorCouch2018},
which does not depend on the presence of a strong magnetic
field. If this asymmetry persists for seconds (and thus much  
longer than the $\sim$0.5\,s over which it was feasible to track
the PNS evolution in recent 3D models), it is
conceivable that LESA could well account for NS kick velocities of
100--200\,km\,s$^{-1}$. It may be speculated whether
this effect could be connected to a constant floor
value of the NS kick velocity of more than 100\,km\,s$^{-1}$ that has
been surmised by \citet{BrayEldridge2016}.
An open question is whether such a
substantial floor value by a neutrino-induced kick would be compatible
with the large population of NSs in globular clusters 
\citep{Katz1975,Katz1983,Pfahletal2002,KuranovPostnov2006} and 
with the dichotomy
of kick velocities advocated by \citet{Katz1975,Pfahletal2002a,
Podsiadlowskietal2004,BeniaminiPiran2016} and \citet{Taurisetal2017},
who invoke a low-kick population of NSs to explain
the existence of a certain class of high-mass X-ray binaries and the
properties of binary NS systems.

Solar-metallicity iron-core progenitors between $\sim$9\,$M_\odot$
and $\sim$12\,$M_\odot$ possess much shallower density profiles
exterior to their iron cores and therefore higher 
core-compactness values than O-Ne-Mg-core progenitors.
Therefore, it is conceivable that some of these cases
might explode with sufficiently large anisotropy of the mass
ejection to account for NS kicks beyond 100\,km\,s$^{-1}$.
Of course, referring to the 
possibility of sufficiently large NS kick velocities in
low-mass iron-core SNe requires confirmation by an extensive
set of 3D simulations of such explosions.
Moreover, a better understanding is needed whether the
explosion properties of SN~1054 (energy, iron production,
light curve) and the chemical composition of the Crab Nebula
are compatible with an origin from a low-mass iron-core
progenitor. 

In summary, our work reveals problems with the widely-used
interpretation of the Crab Nebula as remnant of an ECSN.
The Crab pulsar's proper motion cannot be explained by 
hydrodynamic kicks associated with the ejecta asymmetries
of ECSNe. Considering the pulsar velocity as a consequence
of a natal kick, a connection of Crab to the explosion of a
low-mass iron-core progenitor seems more plausible. Such 
SNe can also have low explosion energies and correspondingly
little nickel production \citep{Jankaetal2012,Melsonetal2015,
Mueller2016,Wanajoetal2018,Radiceetal2017}, compatible with
recent rexamination of the evolution of the Crab
pulsar wind nebula and its interaction with the SN ejecta
\citep{YangChevalier2015},
which corroborates that SN~1054 was a low-energy explosion
($\sim$$10^{50}$\,erg), and that significant energy in 
extended fast material around the Crab is unlikely.
To assess the conflicting progenitor
possibilities, further studies of the Crab Nebula by detailed
observations, in particular also of the chemical composition,
are desirable. Moreover, self-consistent 3D SN models are
needed for a large sample of low-mass (O-Ne-Mg and Fe-core)
progenitors, clarifying the question how steep the increase
of the maximum achievable NS kick velocity with higher
core-compactness can be.

\acknowledgements
\section*{Acknowledgements}
We are grateful to Annop Wongathanarat for providing the 3D Yin-Yang version 
of the \textsc{Prometheus-HOTB} code, to Ken Nomoto for providing the 
stellar progenitor model, to Lorenz H\"udepohl for the post-bounce 
data of the collapsed progenitor star, to Thomas Ertl and Tobias Melson for 
helpful discussions and valuable information, and to Paz Beniamini, 
Jonathan Katz, Bernhard M\"uller, and Alexander Tutukov for
insightful comments on the first arXiv version.
We further thank Elena Erastova (Max Planck Computing and Data Facility)
for the 3D visualization of the simulation shown in Fig.~\ref{fig:elena_plots}.
This research was supported by the Deutsche Forschungsgemeinschaft through
Excellence Cluster Universe (EXC~153; http://www.universe-cluster.de/) and
Sonderforschungsbereich SFB~1258 ``Neutrinos and Dark Matter in Astro- and Particle 
Physics'', and by the European Research Council through grant
ERC-AdG No.\ 341157-COCO2CASA. The computations of the presented models
were carried out on {\em Hydra} of the Max Planck Computing and Data
Facility (MPCDF) Garching.

\software{\textsc{Prometheus-HOTB} 
\citep{Fryxelletal1989, JankaMueller1996, Kifonidisetal2003, Schecketal2006, 
Arconesetal2007,Wongwathanaratetal2013},
VisIt ({\tt https://wci.llnl.gov/simulation/computer-\-codes/\-visit/}).}

\bibliographystyle{apj}
\bibliography{references}

\end{document}